\documentclass[acmlarge,nonacm]{acmart}

\usepackage{algorithm}
\usepackage{amsmath}
\usepackage{amsfonts}
\usepackage{array}
\usepackage{booktabs}
\usepackage{caption}
\usepackage{color}
\usepackage{colortbl}
\usepackage{csquotes}
\usepackage[us,12hr]{datetime}
\usepackage{float}
\usepackage{geometry}
\usepackage{graphicx}
\usepackage{hyperref}
\usepackage{layouts}
\usepackage{longtable}
\usepackage{multirow}
\usepackage{nameref}
\usepackage[percent]{overpic}
\usepackage{pdflscape}
\usepackage{rotating}
\usepackage{tabularx}
\usepackage{textcomp}
\usepackage{url}
\usepackage{wasysym}
\usepackage{xcolor}
\usepackage{xparse}

\makeatletter
\renewenvironment*{displayquote}
{\begingroup\setlength{\leftmargini}{0.3cm}\csq@getcargs{\csq@bdquote{}{}}}
{\csq@edquote\endgroup}
\makeatother

\makeatletter
\let\@authorsaddresses\@empty
\makeatother

\usepackage{url}

\usepackage{etoolbox}
\apptocmd{\sloppy}{\hbadness 10000\relax}{}{}

\begin{document}
\pagestyle{plain}

\title{A Close Look at a Systematic Method\\for Analyzing Sets of Security Advice}

\author{David Barrera}
\affiliation{%
    \institution{Carleton University}
    \country{Canada}
}

\author{Christopher Bellman}
\affiliation{%
    \institution{Carleton University}
    \country{Canada}
}
\authornote{Authors listed in alphabetical order. Contact author: Christopher Bellman (chris@ccsl.carleton.ca). A version of this paper is to appear in Journal of Cybersecurity. Version: 18 May 2023}

\author{Paul C. van Oorschot}
\affiliation{%
    \institution{Carleton University}
    \country{Canada}
}

\begin{abstract}
    We carry out a detailed analysis of the security advice coding method (SAcoding) of Barrera et al. (2021), which is designed to analyze security advice in the sense of measuring actionability and categorizing advice items as practices, policies, principles, or outcomes. The main part of our analysis explores the extent to which a second coder's assignment of codes to advice items agrees with that of a first, for a dataset of 1013 security advice items nominally addressing Internet of Things devices. More broadly, we seek a deeper understanding of the soundness and utility of the SAcoding method, and the degree to which it meets the design goal of reducing subjectivity in assigning codes to security advice items. Our analysis results in suggestions for modifications to the coding tree methodology, and some recommendations. We believe the coding tree approach may be of interest for analysis of qualitative data beyond security advice datasets alone.
\end{abstract}


\maketitle

\section{Introduction}

There is no shortage of security advice, in a wide range of domains.  With the rise in popularity of the Internet of Things (IoT), and a corresponding rise in consumer IoT devices with security vulnerabilities \cite{Alrawi2019, Kolias2017}  numerous organizations have offered security advice positioned as IoT security guidelines, recommendations, baseline requirements, best practices, and codes of practice (e.g.,  \cite{ETSI2020, DCMS2}). Barrera et al.\ \cite{Barrera2021a} recently proposed a method called security advice coding (SAcoding, next section) to analyze such advice datasets. They used a so-called \textit{coding tree} and a single coder to analyze what is referred to as the cb1013-dataset~\cite{BellmanDataset}, resulting from filtering out identical items from a UK government compilation of 1052 advice items \cite{IoTSecMap}. Bellman and van Oorschot~\cite{bellman2022a} used the same method to analyze and compare two much smaller and coarser (higher-level) advice datasets, again primarily using a single coder (with light review from a second). The main focus of the present paper is a critique the SAcoding method itself (rather than introducing it or demonstrating its use), through comparing the results from a second coder, on the same large dataset, to those from the single coder in the original paper \cite{Barrera2021a}. The analysis methodology, including insights gained from what we call \textit{tag-vs-tag tables}, aids exploration of the original method, and suggests opportunities to improve it.

Our analysis of the SAcoding method is motivated by a desire to increase our understanding of it, its utility, and what it can be relied upon to deliver. We set out to identify which of its aspects deserve our confidence, and where or how it might be improved. One improvement direction would be any changes that increase consistency, between different coders, in codes assigned to advice items (\textit{tag agreements}). From our analysis, we note limitations and offer insights and recommendations on the structure of the coding tree, instructions to coders, and the descriptions associated with questions at tree nodes. Collectively, this may encourage others in the community to use SAcoding to analyze other sets of security advice, independently critique or improve the method, and perhaps create and design alternative systematic methods for analyzing security advice.

\section{Background (on SAcoding method)}

Qualitative data \textit{coding} \cite{Corbin2008} is commonly used to manually extract themes and perform analysis of qualitative data such as verbalized thoughts and interview transcripts from study participants. It is an imprecise process by which review of a dataset results in development of a \textit{codebook} containing refined labels for concepts derived from the data. In a perfect world, after a codebook is built, independent coders would code given data items identically; in practice, the process is subjective. Ideally, different coders applying the SAcoding method to security advice items would produce the same code for each item (perfect between-coder agreement), and an individual coder would duplicate their results on repeated uses (within-coder agreement). While this rarely occurs in practice (as we later show), our analysis herein explores the degree of between-coder agreement for two coders. We also believe that consistency in this sense, which may be less important in typical qualitative coding where the goal is to extract themes or sentiments, is perhaps more important in the context of SAcoding where the resulting codes are used to analyze and critique specific security advice items; here inconsistency undermines the reliability of an analysis, e.g., in distinguishing actionable from non-actionable advice.

The SAcoding method \cite{Barrera2021a} guides analysts (coders) through a set of questions to ultimately assign a pre-defined category code or \textit{tag} per Table~\ref{box-now-table:tags} (e.g., outcome, principle, policy, or one of several categories of practices) to individual security advice items. Tags $P_3$--$P_6$ are used for practices that are pre-classified as \textit{actionable}, meaning they specify step-by-step instructions for advice followers, whereas, e.g., tag $T$ (\textit{Desired Outcome}) is for an advice item that states security outcomes but suggests no means to reach them; $P_3$ (\textit{Infeasible Practice}) is for practices that are actionable, but for which advice targets might not have sufficient resources to carry out the advice.

\begin{table}[!tb]
    \centering
    \caption{SAcoding tree codes, from Barrera et al.\ \protect\cite{Barrera2021a} with some descriptions reworded. The two distinct codes for \textit{Desired Outcome} allow tracking of different code tree paths. The single-quoted `Practice' in tag names for $P_1$, $P_2$ signal that these are not actionable practices. *denotes codes considered actionable.} 
    \label{box-now-table:tags}
    {\small
    \begin{tabular}{@{} p{0.9cm} @{}  p{4.5cm} @{} p{8.5cm}@{} }
        Code &  Tag name & Description \\ \toprule
        
        $M_1$ & Not Useful (vague/unclear 
        \newline $~~~~~~~~~~~~~~~~~$ or multiple items) &    Advice that doesn't make sense linguistically (e.g., unclear grammar), or isn't focused on a specific task/action. An item tagged $M_1$ may optionally be given a supplementary label, \textit{Unfocused}. \\
        
        $M_2$ & Beyond Scope of Security & Advice that is hard to argue could possibly help security. \\
        
        $N$ & Security Principle & Advice in the form of a general rule (broadly applicable), that historically improves security outcomes, or reduces exposures. \\
        
        $P_1$ & Incompletely Specified `Practice' & Advice appearing at first glance to give technical direction like a practice (e.g., technical mechanism, specific rule), but lacking clear indication of steps to take. The tag is thus considered `non-actionable'. \\
        
        $P_2$ & General Policy (General `Practice') & Advice that indicates a general approach (or general policy) but gives no explicit techniques or tools. The tag is considered non-actionable (despite its name) due to its general, unspecific nature.  \\
        
        *$P_3$ & Infeasible Practice & A practice that would consume unreasonable resources (time, money), thereby failing most/all cost-benefit analyses.  \\
        
        *$P_4$ & Specific Practice/Security Expert & A practice requiring the expertise/deep knowledge of a security expert; may require inferring steps that are not explicit in the advice. \\
        
        *$P_5$ & Specific Practice/IT Specialist & A practice executable by typical IT workers (but beyond typical end-users), using basic professional knowledge of security. \\
        
        *$P_6$ & Specific Practice/End-User & A practice executable by typical end-users, e.g., by directly interacting with a device, mobile app, or cloud service. \\
        
        $T$, $T'~~$ & Desired Outcome & Advice suggesting a generic, high-level end goal to attain, but lacking a specific method by which to reach that outcome. \\
        \bottomrule
    \end{tabular}
}
\end{table}

\begin{figure}[t]
    \centering
    \includegraphics[width=0.90\textwidth]{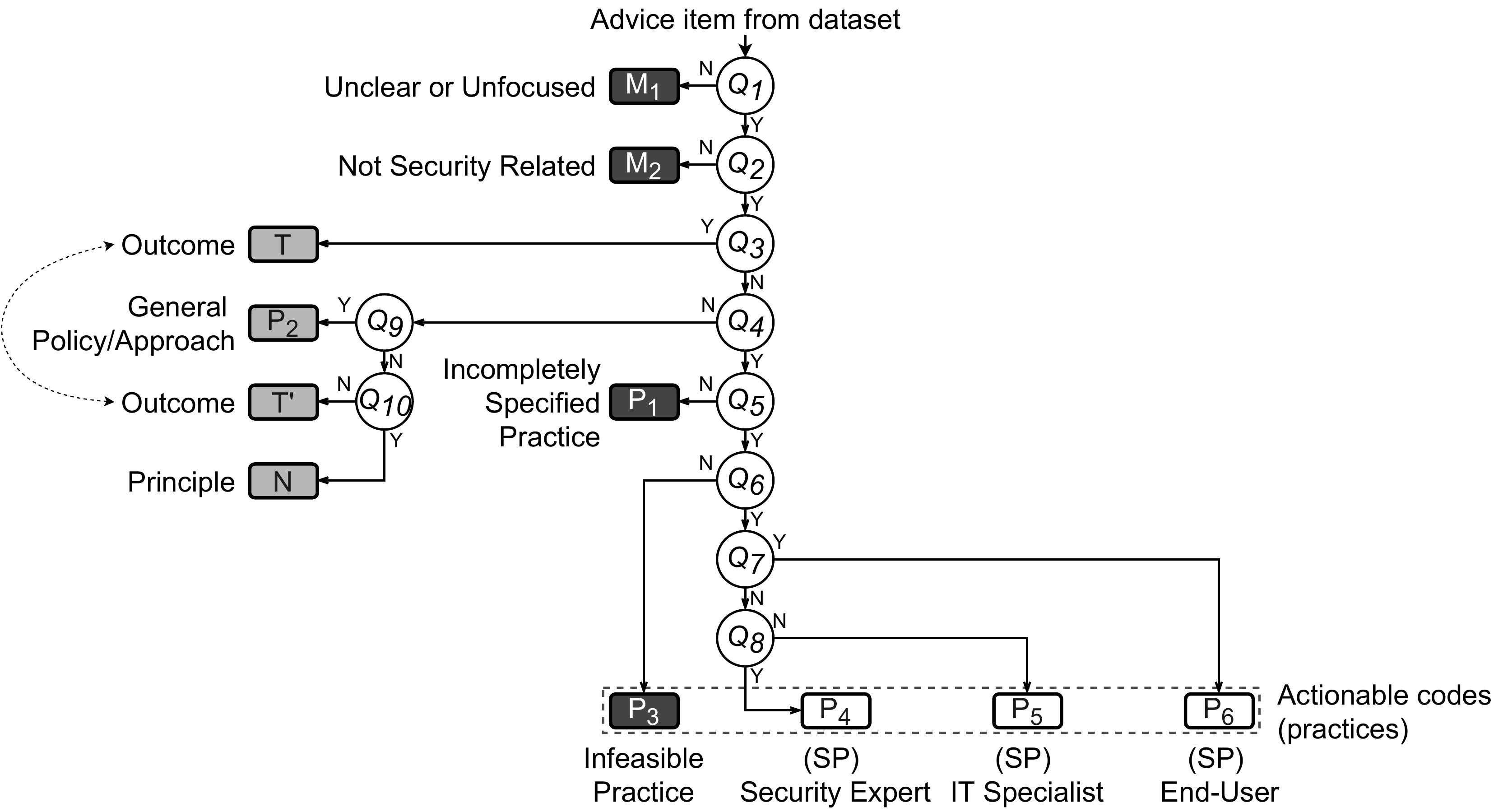}
    \vspace{-10pt}
    \caption{SAcoding tree, from Barrera et al.\ \protect\cite{Barrera2021a} with minor modifications. Answering questions $Q_i$ (Table~\ref{nowTable:treeQuestions}) leads to advice items being assigned leaf node codes. Codes are described in Table~\ref{box-now-table:tags}. Black tags indicate advice items that may benefit most from revisiting by advice givers. \textit{SP} is short for \textit{Specific Practice}.}
    \label{fig:flowchart2}
    \vspace{-5pt}
\end{figure}

The method uses a top-rooted \textit{coding tree} (Fig.~\ref{fig:flowchart2}), with leaf nodes holding the tags. Each interior node has one \textit{yes}/\textit{no} question. For a given advice item, the root question is asked. The answer dictates whether the \textit{yes} or \textit{no} branch is followed to the next node (question). Reaching a leaf results in the leaf's code being assigned to the advice item. By the existing method, a coder can apply up to two tags (but at most once per item) to an advice item if they can justify both a \textit{yes} and \textit{no} answer at a question node. In this case, two tags result and both are considered equally valid for the advice item \cite{Barrera2021a}.

\begin{table}[b]
    \centering
    \vspace{-5pt}
    \caption{SAcoding tree questions, asked for each advice item. For supplementary annotations available to coders (providing additional context), see Barrera et al. \protect\cite{Barrera2021a}, which is also the source of the questions. Question $Q_{11}$ omitted (as discussed later).}
    \label{nowTable:treeQuestions}
    {\small
        \begin{tabular}{p{0.9cm} @{} p{10.0cm}}
        No.\ & Question (yes/no answer determines next coding tree question asked) \\ \toprule
        $Q_1$ & Is the item conveyed in unambiguous language, and relatively focused? \\ 
        $Q_2$ & Is it arguably helpful for security? \\ 
        $Q_3$ & Is it focused more on a desired outcome than how to achieve it? \\ 
        $Q_4$ & Does it suggest a security technique, mechanism, software tool, or specific rule? \\ 
        $Q_5$ & Does it describe or imply steps or explicit actions to take? \\ 
        $Q_6$ & Is it viable to accomplish with reasonable resources? \\ 
        $Q_7$ & Is it intended that the end-user carry out this item? \\ 
        $Q_8$ & Is it intended that a security expert carry out this item? \\ 
        $Q_9$ & Is it a general policy, general practice, or general procedure? \\ 
        $Q_{10}$ & Is it a broad approach or security property?  \\ 
        \bottomrule
        \end{tabular}
    }
\end{table}

The publication that introduced SAcoding \cite{Barrera2021a} did not explicitly recommend the number of coders to use, but employed the method using a single coder on a large dataset. Ideally, the SAcoding tree would produce very few or zero tag nonagreements as noted above, with each question in the tree designed to reduce reliance on subjective interpretation by coders---for a given advice item, every question having crisp, clear criteria, such that competent coders all answer \textit{yes} or \textit{no} in unison. Our main goal herein is to explore the degree to which this occurs, and we find that coder nonagreements occur more often than we implicitly expected (as designers of the original method), leading us to suggest modifications to the method.

In our analysis, we use the previously mentioned cb1013-dataset \cite{BellmanDataset}. As one aspect, we look to identify coding tree questions at which a disproportionately large (or small) number of nonagreements occur, and consider potential underlying reasons. 

\section{Methodology and Results of Coding}

The previous section reviewed the SAcoding method. Here we discuss our use of it to tag the full 1013-item dataset~\cite{BellmanDataset} as a second coder (C2), and compare results with coding of the same dataset by a coder we call C1 from previous work \cite{Barrera2021a}. We use the same procedures as C1 to produce the C2 results (same tree, coder instructions, coding interface software). Our methodology is compared to others later, under Related Work.

\begin{figure*}[t]
    \centering
    \includegraphics[width=0.95\textwidth]{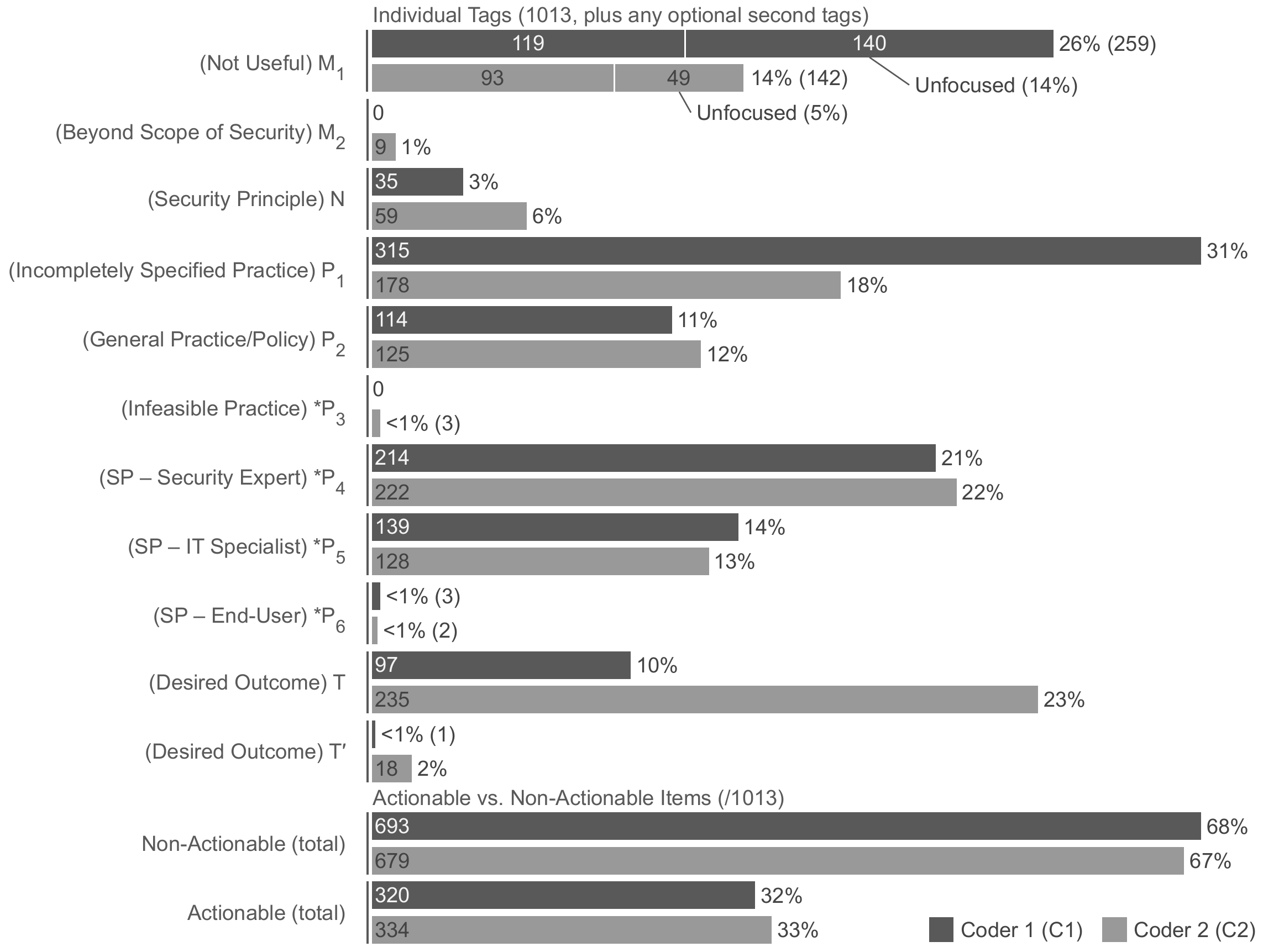}
    \vspace{-10pt}
    \caption{Tag distribution from coding the 1013-item dataset. C1's results duplicated from \cite{Barrera2021a} to allow comparison. Summing from figure, total tag counts for C1 (1013+164) and C2 (1013+108) exceed 1013 due to optional second codes. Actionable/Non-Actionable bars use separate scale. Note: while the graph shows the number of tags of a given type for each coder, identical counts for, say, \textit{$P_5$} would not imply the same individual items were identically tagged \textit{$P_5$} by both coders (see later section `Deeper View'). *denotes actionable codes.}
    \label{fig:tagresults2}
\end{figure*}

After C2 finished tagging the 1013 items, we compared the set of tags from C2 to that from C1; the coding results from both coders are publicly available in a file that integrates the 1013-item dataset \cite{BellmanDatasetJCS}. We first explore differences in tag frequency and distribution, to get a sense of the tags each coder applied to the advice item dataset. For each coding, we first separately summarize the number of items tagged in each tag category. In the tag set for each coder, each of the 1013 advice items was tagged with  one or two tags. Fig.~\ref{fig:tagresults2} summarizes the counts; due to optional second tags, the per-coder sum of all tag counts exceeds 1013. We later give a more detailed analysis (starting with Table~\ref{tab:c1vsc2}) showing tag distributions for C1 and C2 partitioned across 13 advice-item groups as explained later. 

As an aside, Barrera et al.~\cite{Barrera2021a} note that the coding software interface they used (also used herein) did not prevent coders from ``short-cutting'' by directly assigning a tag to an item (bypassing the method); they suggest that any software interface tool used should preclude this.  This change was not made in the tool we used, but inquiries with both the original coder C1 \cite{Barrera2021a} and our coder C2 confirmed that short-cutting was generally avoided. 
    
\subsection{Proportions of Actionable Advice and Individual Actionable Tags}

An advice item is considered actionable (for a given coder) if either the first or (when present) second tag assigned to it is actionable (one of \textit{$P_3$}--\textit{$P_6$}, per Table \ref{box-now-table:tags}); the justification is that the coder has  an argument supporting actionability. Otherwise, the item is considered non-actionable. The Actionable and Non-Actionable bars of Fig.~\ref{fig:tagresults2} show the resulting counts for each coder. The count across these two bars thus sums to 1013 for each coder. 

Fig.~\ref{fig:tagresults2} shows that 33\% of items received an actionable tag from C2, a proportion in line with the 32\% proportion of actionable items in C1's earlier analysis \cite{Barrera2021a} of this 1013-item dataset. (The same individual items, however, did not necessarily receive the same tags.) Further, C1 and C2 agreed on whether a given advice item was actionable (actionable for both, \textit{or} non-actionable for both) in 80\% of individual items (Table~\ref{tab:disagreementsummary}, discussed later); we repeat, this differs from comparing overall proportions of the dataset that each tagged as actionable (Fig.~\ref{fig:tagresults2}). As such, while one might claim that this supports the assertion that the coding tree---its questions, their ordering, and wording---enables effectively classifying items as actionable vs.\ non-actionable in terms of overall proportion, it does not support the claim that the method allows effective such classification of individual items. 

Using Fig.~\ref{fig:tagresults2}, we now consider individual actionable tags ($P_3$--$P_6$). Out of $2 \times 1013$ opportunities, \textit{$P_6$} (advice for \textit{End-User}) was used only 5 times, and \textit{$P_3$} (\textit{Infeasible}) only 3 times. The $P_3$ result suggests that advice in this dataset is largely feasible, in the case that it is actionable; the $P_6$ result suggests that this advice set does not target end-users. The low use of \textit{$P_3$} and \textit{$P_6$} is discussed further under `Low Numbers of Q-nonagreements'. 

Next consider $P_4$ (advice requiring knowledge of a \textit{Security Expert}) and $P_5$ (requiring knowledge of an \textit{IT Specialist}). C1 and C2 used comparable proportions of \textit{$P_4$} (21\% vs 22\%), and likewise for \textit{$P_5$} (14\% vs 13\%), but we see $P_4$ used 54\% and 73\% more, resp., than $P_5$. Recalling that the primary target of this 1013-item advice dataset appears to be  IoT device manufacturers (or pre-deployment stakeholders) \cite{DCMS1, DCMS2, ETSI2020}, we find this target consistent with the healthy proportions of items tagged $P_5$, and low proportions of $P_6$. However, the relatively high proportions of $P_4$ appear inconsistent with this dataset being actionable by IoT device manufacturers, whom we expect (at least for small and medium-sized companies) do not always employ security experts. 
    
\textbf{Proportion gaps in non-actionable tags.} 
For this dataset, while both coders assigned similar numbers of tags in each actionable tag category,  this was not so for several non-actionable tags, including: 
    
\begin{itemize}
    \item \textit{$M_1$} (\textit{Unclear or Unfocused}): $26$\% vs. $14$\% (for C1   and C2, respectively) 
    
    \item \textit{$P_1$} (\textit{Incompletely Specified Practice}): $31$\% vs. $18$\%
    
    \item \textit{$T$ or $T'$} (\textit{Desired Outcome}): $10$\% vs. $25$\%  
    
    \item \textit{$N$} (\textit{Security Principle}): $3$\% vs. $6$\%
\end{itemize}

\noindent For the other 6 code categories, coders had similar tag distributions in that the number of items assigned each tag differed between coders by at most $1$ per cent of the number of items in the dataset. The latter is promising, but overall it appears premature to claim that the SAcoding method can reliably estimate the proportion of a dataset's advice items in (most) categories; while plausible for some datasets and some coders,\footnote{Here our results are for one dataset and two coders, with a further limitation that the coders are also authors and thus not at arms-length.} further evidence and perhaps methodology changes (discussed later) would be needed to support a claim that tag distribution is reproducible.\footnote{
    Following ACM definitions \cite{ACMTerminology}, \textit{repeatability} refers to obtaining the same results with the same coders using the same experimental setup (SAcoding method, interface tool, and dataset); \textit{reproducibility} obtains the same results for different coders and the same experimental setup; and \textit{replicability} means the same results are obtained for different coders using a different setup (e.g., independently created interface tool, but same dataset).
} We offer no generic metric for what might be considered an acceptable level of agreement (consistency); we discuss the need to investigate measuring reproducibility beyond two coders under \textit{Limitations and Recommendations}. 

As noted earlier, similar cross-coder proportions of actionable vs.\ non-actionable tags does not imply agreements about the actionability of individual items; the same is true for specific tag values (as opposed to their actionability). For example, the proportions of the specific tag value $P_5$ in Fig.~\ref{fig:tagresults2} (14\% and 13\%) provide no information on how often C1 and C2 tagged the same individual advice items with $P_5$; from tag-vs-tag details (Tables~\ref{tab:heatmap_2tag} and \ref{tab:heatmapSD_3tag} given later), we see that of Fig.~\ref{fig:tagresults2}'s $139$ $P_5$ tags for C1, $43+29$ resulted in nonagreement, and similarly of $128$ for C2, $47+37$ resulted in nonagreement, thus overall $(72 + 84)/(139 + 128) = 58.4$\% of $P_5$ tags yielded nonagreements, despite similar proportions. This reveals a discrepancy between coder tag \textit{proportions} and between-coder tag \textit{agreement}, and a disappointing level of same-item coder agreement.

As the second comment, of the three large percentage gaps noted above, the gap for \textit{T} \textit{(Outcome)} is perhaps of greatest concern, as differentiating outcomes from actionable practices was a goal~\cite{Barrera2021a}.  However, this concern is partially mitigated by data from later Table~\ref{tab:heatmap_2tag}, showing that most tag nonagreements involving $T$ are with other non-actionable tags (primarily $P_1$, $M_1$, $P_2$).

Finally, further discussion of nonagreements involving $M_1$ is given later under `Question 1', and similarly for $T$ under `Question 3' and in discussion of the tag-vs-tag tables. 

\subsection{Terminology for Multi-Coder Agreements and Nonagreements}

We next consider at which questions coders had \textit{agreements} and \textit{nonagreements} for each advice item. We use agreements and nonagreements in two contexts. An \textit{agreement} in the context of \textit{tags} (\textit{T-agreement}) is when at least one of the tags given to an advice item by a coder is the same as one of the tags given by the other coder; otherwise, it is a \textit{T-nonagreement} (the coders have no tags in common among their first and possibly second tags for the advice item). In the context of \textit{a question}, an agreement (\textit{Q-agreement}) is when both coders provide the same answer to a question in the tree (both answer \textit{yes}, or both answer \textit{no}), for a given advice item. For a given tree question, when one coder answers \textit{yes} and the other \textit{no}, we say it is a \textit{Q-nonagreement}. 

Clearly, Q-agreements are related to T-agreements, e.g., each coder assigning the same code through the tree implies Q-agreements at all questions on the path to that code (leaf). A T-agreement is determined based on tags assigned to advice items through use of the coding tree; Q-agreements are determined based on individual questions in the coding tree that both coders answer for the same advice item.

How tag and question agreements and nonagreements are determined is described more thoroughly below. We describe non-matching tags and question answers as nonagreements instead of disagreements, based on our view that the term \textit{disagreement} implies two coders explicitly disagreed on something (e.g., the tag to be applied, or the answer to a question), rather than coders independently selecting different sequences through the coding tree and the outcomes differing. Understanding where coders had nonagreements most frequently (e.g., which questions in the coding tree they diverge on) allows us to consider where the coding tree might be improved to more reliably converge coder answers toward \textit{yes} or \textit{no}, and to identify advice items that may be vague or open to subjective interpretation by a coder. 

A \textit{coding sequence} (hereafter \textit{sequence}) is the sequence of question nodes resulting from use of the coding tree on an advice item; the nodes are joined by edges, as determined by \textit{yes}/\textit{no} answers (in Fig.\ref{fig:flowchart2}, the edges labeled \textit{Y}/\textit{N} from $Q_1$ to a leaf code). The number $n$ of nodes in a sequence is the number of questions a coder answers before reaching a tag. For example, tag \textit{$P_1$} is reached by answering Questions 1 (\textit{yes}), 2 (\textit{yes}), 3 (\textit{no}), 4 (\textit{yes}), and 5 (\textit{no}), yielding sequence $(Q_1, Q_2, Q_3, Q_4, Q_5)$ with $n=5$. As a sequence describes the path through the coding tree to reach a single code, a coder determines one sequence for each tag assigned to an item. Sequences are indirectly created by coders, as a result of answering a question at each node until reaching a tag. 

For a given advice item, a \textit{diverging question} is the tree question at which two coders give different answers, thus taking a different exit path from that node, eventually yielding different tags. When two coders yield a different tag on a given advice item, the diverging question can be determined by tracing backwards up the tree, from each of the two tag leaf nodes (Fig.~\ref{fig:flowchart2}), until these paths intersect at a question node. 

As the number of tags given to an advice item by two coders can vary from 2 to 4, we consider three \textit{types} of advice tag comparison (based on T-agreements). Each comparison is within the scope of a single advice item, i.e., by looking at the tags each coder (C1 and C2) assigned to that item. As first and second tags (resp.\ denoted by subscripts, e.g., $C1_1$ and $C1_2$) are given equal importance \cite{Barrera2021a}, these types are based on the \textit{number} of tags assigned by coders, not which tags were assigned first or second. In all three comparison types, a \textit{match} implies two identical tags (e.g., $C1_1 = C2_1$); one tag must be from each coder. 
 
\textbf{SS-type Tag Comparisons.} 
In an \textit{SS-type} comparison (single-single), each coder gave only one tag to the item, and a T-agreement occurs if the tags match, i.e., $C1_1 = C2_1$; a T-nonagreement occurs otherwise. 

\textbf{SD-type Tag Comparisons.} 
In an \textit{SD-type} comparison (single-double, including also double-single), one coder gave one tag (first), and the other opted for two tags (first and second), for a total of three tags to one advice item. In this type, a T-agreement occurs if the single-tag coder's first tag matched either the double-tag coder's first \textit{or} second tag (recall that the first and second tags are considered to be of equal importance); a T-nonagreement means the single-tag coder's tag was identical to neither of the double-tag coder's tags. 

In SD-type comparisons, to determine a single question where coders diverged (a Q-nonagreement occurred), we use the longest T-nonagreement overlapping sequence (of the two comparisons, e.g. \{$C1_1$, $C2_1$\} and \{$C1_1$, $C2_2$\}), and declare the final question in that sequence to be the diverging question. If we used the shortest of the two overlapping sequences, in every instance that a coder tagged an item as \textit{$M_1$} ($Q_1$'s \textit{no} answer), the diverging question would be $Q_1$, as it is the final question in the overlapping sequence.\footnote{
    This case is discussed further under `High Numbers of Q-nonagreements (Question 1)'. 
} As in a T-nonagreement at most one coder could yield an \textit{$M_1$} tag (otherwise a T-agreement results), the other sequence of the two-tag coder is necessarily longer, providing a longer overlapping sequence to analyze. We observed no cases where one advice item had two overlapping sequences of the same length, with different leaf tags reached. 

\textbf{DD-type Tag Comparisons.} 
In a \textit{DD-type} comparison (double-double), both coders opt for two tags (first and second), resulting in a total of four tags assigned to an advice item. In this case, we declare a T-agreement if one coder's first or second tag is identical to either tag of the other coder; a T-nonagreement occurs if neither of the two tags from the first coder is identical to either from the second coder. As so few instances occurred where both coders used two tags (as discussed shortly), type DD is excluded from most of our subsequent analysis; however, we summarize tag results for the 19 relevant items (cf. Table~\ref{tab:disagreementsummary}) in Table~\ref{tab:DDitems}.

\begingroup
\setlength{\tabcolsep}{4pt}
\begin{table}[tb!]
    \centering
    \caption{Codes assigned by each coder for the 19 items resulting in DD-type comparisons. Column headers give the index of each relevant advice item in the dataset (\textit{cb1013-dataset-twocoder} \cite{BellmanDatasetJCS}).  While these results are not interpreted herein, they are noted explicitly to allow independent exploration of advice items that resulted in DD-type comparisons.}
    \label{tab:DDitems}
    \small{
    \begin{tabular}{@{}cccccccccccccccccccc@{}}
        Coder & \multicolumn{19}{c}{Index of advice item in dataset} \\ \cline{2-20}
        tag & 24 & 67 & 130 & 165 & 292 & 317 & 324 & 404 & 465 & 519 & 534 & 543 & 551 & 680 & 686 & 691 & 734 & 787 & 801 \\ 
        \midrule
        $C1_1$ & $P_5$ & T & $P_4$ & $P_4$ & $P_4$ & $P_4$ & $P_4$ & $P_4$ & N & $P_2$ & $P_5$ & $P_2$ & $P_1$ & $P_1$ & $P_1$ & $P_4$ & $P_5$ & $P_4$ & $P_1$ \\
        $C1_2$ & $P_4$ & $P_5$ & $P_1$ & $P_1$ & $P_5$ & $P_1$ & $P_5$ & $P_5$ & $M_1$ & $P_1$ & $P_4$ & $P_1$ & $P_4$ & $P_4$ & $P_4$ & $P_1$ & $P_4$ & $P_1$ & $P_4$  \\ \midrule
        $C2_1$ & $P_2$ & $P_1$ & T & $P_5$ & $P_1$ & $P_1$ & $P_4$ & $P_5$ & N & T & $P_1$ & $P_2$ & $P_4$ & $P_5$ & $P_5$ & $P_1$ & $P_1$ & $P_4$ & $P_1$ \\
        $C2_2$ & $P_1$ & $P_4$ & $P_1$ & $P_4$ & $P_4$ & $P_4$ & $P_1$ & $P_4$ & $M_1$ & $P_2$ & $P_4$ & T & $P_1$ & $P_4$ & $P_4$ & $P_4$ & $P_4$ & $P_1$ & $P_4$ \\ 
        \bottomrule
    \end{tabular}
    }
\end{table}
\endgroup

\subsection{T-agreement Summary and Results} 

Table~\ref{tab:disagreementsummary} summarizes the results of calculating T-agreements (type SS, SD, and DD) as discussed. 760 advice items received one code from each coder (SS-type), 234 had one code from one coder and two from the other (SD-type), and 19 had two codes from each coder (DD-type). The high percentage of T-agreements among DD-type advice items are not surprising, based on the larger number of pairs available for a match due to a second tag from both coders (Table~\ref{tab:disagreementsummary}'s third column shows the proportion of T-agreements increasing with the number of tags in the comparison set). Type SS and SD comparisons account for 98\% of two-coder tag comparisons (column 2). As only 2\% of advice items were tagged twice by both coders, we do not analyze or discuss DD-type agreements (of tags or questions) further, beyond Tables~\ref{tab:disagreementsummary} and \ref{tab:disagreementsecondtag}.

\begin{table}
    \centering
    \footnotesize{
    \caption{Number of advice items partitioned by tag comparison type (SS, SD, DD), number of T-agreements resulting from each, and number of actionability agreements (coder tags agree, i.e., both actionable, or both non-actionable). T-nonagreement percentages can be computed as: $100\%$ minus (column 3 percentage). Note column 2 has sum: $760 + 234 + 19 = 1013$.}
    \label{tab:disagreementsummary}
    \begin{tabular}{rrrr}
        Comparison & Number of advice items
        & Number of
        & Actionability \\
        type & 
        for this type (of 1013)& 
        T-agreements
        & 
        agreements \\
        \toprule
        SS & 
        760 (75\%) & 
        315 (41\% of 760) & 
        608 (80\% of 760)\\
        SD & 
        234 (23\%) & 
        130 (56\% of 234) & 
        204 (87\% of 234)\\
        DD & 
        19 $~$ (2\%) & 
        17 $~$ (89\% of 19) & 
        18 (95\% of 19)\\
        \bottomrule
    \end{tabular}
    \vspace{5pt}
    }
\end{table}

\begin{table}
    \centering
    \footnotesize{
    \caption{Partitioning of T-agreements by type and ordered code pairing. Cases consider whether the first or second tag of coders C1, C2 resp., result in T-agreement. 
    T-agreement counts from  Table~\ref{tab:disagreementsummary} (column 3). 
    First-Second includes Second-First.}
    \label{tab:disagreementsecondtag}
    \begin{tabular}{rrrrr}
        Comparison & Number of & \multicolumn{3}{c}{Number (proportion) by ordered code pairing} \\ \cline{3-5}
        type & T-agreements & 
        First-First & 
        First-Second & 
        Second-Second\\
        \toprule
        SS & 315 & 
        315 (100\%) & 
        -- & 
        -- \\
        SD & 130 & 
        78$~$ (60\%) & 
        52 (40\%) & 
        -- \\
        DD & 17  & 
        5$~$ (29\%) & 
        7 (42\%) & 
        5 (29\%) \\
        \bottomrule
    \end{tabular}
    \vspace{-5pt}
    }
\end{table}

Table~\ref{tab:disagreementsecondtag} splits out the T-agreements from Table~\ref{tab:disagreementsummary}, based on tag selection order in the case of optional second tags (\textit{first} and \textit{second} codes), across types SS, SD, and DD. Although (as noted earlier) the analysis views first and second codes as equivalent, this data is included for completeness and allows a check for anomalies.

\subsection{Q-nonagreement Results: Distribution across Questions, and Proportion within Each Question} 

\begin{figure*}[t!]
    \centering
    \includegraphics[width=1.0\textwidth]{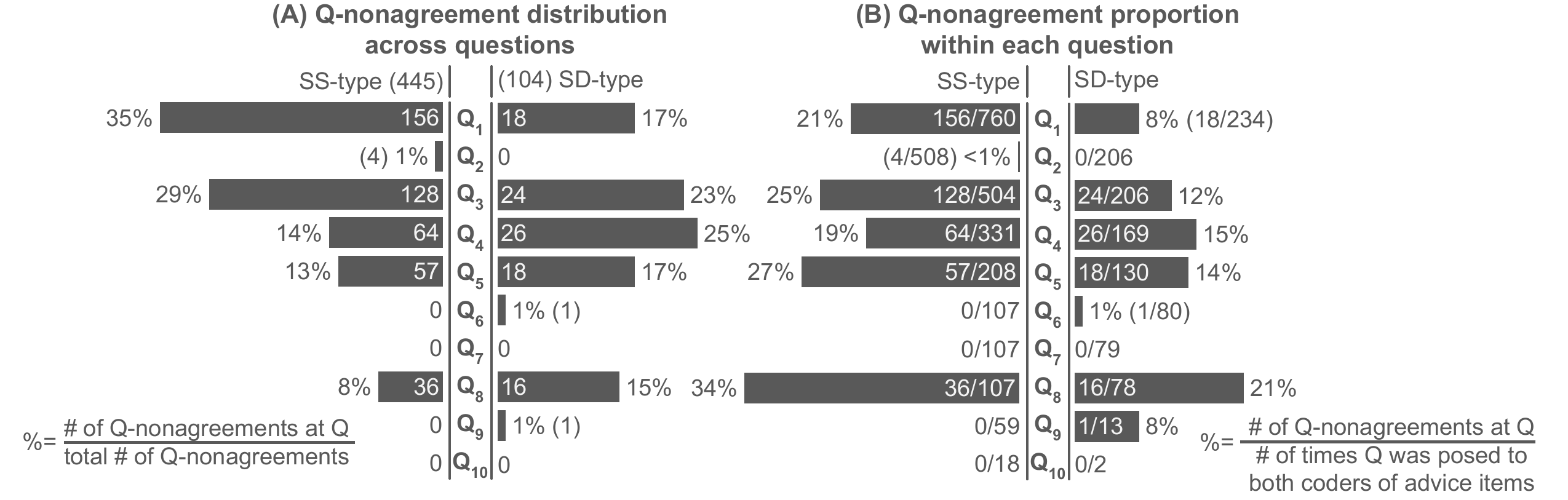}
    \vspace{-15pt}
    \caption{A) Q-nonagreement distribution across questions for two coders' tags. B) Q-nonagreement proportions within each question (showing ratio of: number of Q-nonagreements at a node, to how often both coders encountered that node including both Q-agreements and Q-nonagreements). Fig.~\ref{fig:disagreementtree_both} explains calculation of part B) Q-nonagreements at each question. Values match Table~\ref{tab:disagreementsummary}'s 315 SS-type T-agreements ($760 - 315 = 445$ nonagreements); likewise for SD-type ($234 - 130 = 104$ nonagreements).}
    \label{fig:disagreementbar}
\end{figure*}

\begin{figure*}[!b]
    \centering
    \includegraphics[width=0.65\textwidth]{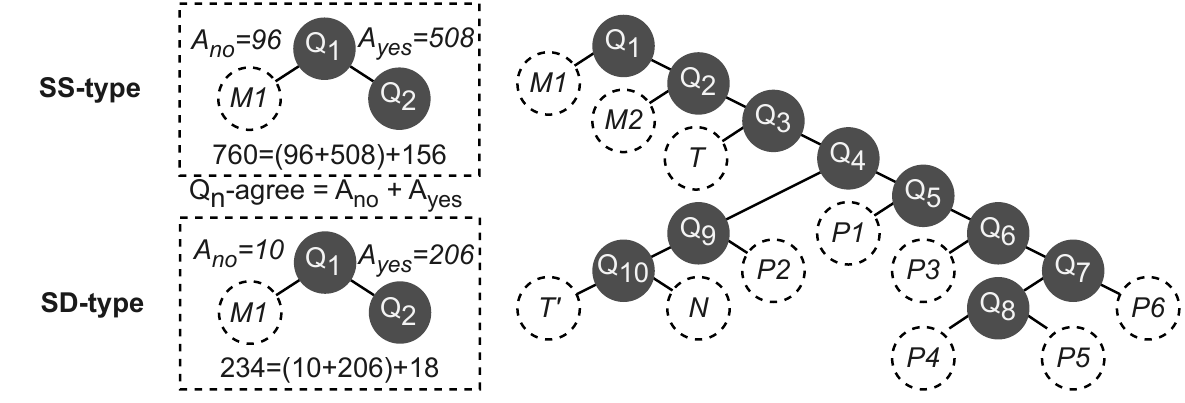}
    \vspace{-10pt}
    \caption{Sum of Q-agreements at each question node (cf.\ Fig.~\ref{fig:disagreementbar}B). Sum includes agreements on both \textit{yes} and \textit{no} answers (see dashed boxes). Number of comparison instances for $Q_1$, from Table~\ref{tab:disagreementsummary}, is: 760 for SS-type (one tag per coder), 234 for SD-type (one coder giving two tags). The miniature tree (right) allows a visual cross-check of how many codes are impacted by decisions at a question.}
    \label{fig:disagreementtree_both}
    \vspace{-5pt}
\end{figure*}

Fig.~\ref{fig:disagreementbar}A gives the number of Q-nonagreements at each question, and their distribution across questions, i.e., the proportion of Q-nonagreements at each question among the total number of Q-nonagreements across all questions. Fig.~\ref{fig:disagreementbar}B shows this same number of Q-nonagreements at each question but now relative to the number of times that question was visited by both coders on the same advice item, i.e., for each question, the percentage of joint visits that yielded Q-nonagreements (see also Fig~\ref{fig:disagreementtree_both}).

To calculate these statistics, we used the longest overlaps between the coders' node sequences, as explained earlier. For each tree question we counted how often coders had a Q-agreement (same answer, both \textit{yes} or both \textit{no}), and the number of Q-nonagreements. When coders agreed on the final tag for an item, the Q-agreement count of each node in the sequence (including final $Q$ node) is incremented. When coders did not agree on the final tag (a $T$-nonagreement), the Q-agreement count was incremented for each node in the overlap sequence excluding the diverging node; and the Q-nonagreement count was incremented for the final node ($Q_n$) in the overlap sequence, as this was the diverging question. Since the first question $Q_1$ is asked regardless of any other nodes visited by each coder, the number of times $Q_1$ is asked is the total number of items in each comparison type (from Table~\ref{tab:disagreementsummary}: 760 for SS-type, 234 for SD-type). This explains how Fig.~\ref{fig:disagreementbar}B was constructed. 

\section{Interpretation of Q-nonagreement Results}

We now interpret the Q-nonagreement results. We examine aspects of the coding tree associated with a disproportionately large number of Q-nonagreements (areas where coding of items differed most across coders), and also small numbers thereof. As discussed later, these are candidate areas to consider for coding tree improvements; questions with very few Q-nonagreements are candidates for removal to simplify the tree. In some cases we provide concrete explanations for why Q-nonagreements take place more or less at some questions; in others, we offer conjectures with less confidence, keeping in mind the subjective nature of qualitative coding. 

\subsection{High Numbers of Q-nonagreements}

As Fig.~\ref{fig:disagreementbar}A shows, for SS-type and SD-type comparisons, nearly all Q-nonagreements (98\% combined) come from Questions 1, 3, 4, 5, and 8. We examine the top three, and consider why these dominate the Q-nonagreements.

\textbf{Question 1} (35\% and 17\% Q-nonagreements for type SS and SD, respectively). $Q_1$ asks if advice is in \textit{unambiguous language} and \textit{relatively focused}. As a possible reason why there is  a greater proportion of Q-nonagreements at $Q_1$ than at other questions, we hypothesize that the wording of $Q_1$ may be interpreted differently by coders C1 and C2, due to differences in their security experience (as discussed under `Limitations') or other differences in personal interpretation independent of experience. For example, some wording in an advice item may appear to be \textit{unambiguous language} to one coder but not another. Unclear wording of the question itself such as this term and the second key term in $Q_1$, \textit{relatively focused} (intending to ask whether sub-items are focused on the same topic, i.e., related), may contribute to Q-nonagreements. Here a \textit{sub-item} is part of a larger integrated item that may cover sub-topics that are distinct, or only loosely-connected, or covering multiple aspects of a single or closely-related topics; sub-items appeared often in this dataset \cite{Barrera2021a}.

\begin{figure*}[!b]
    \centering
    \includegraphics[width=\textwidth]{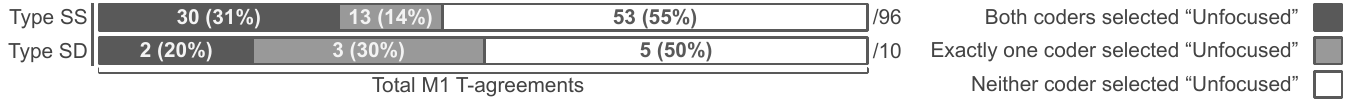}
    \vspace{-15pt}
    \caption{Examination of \textit{Unfocused} sub-label in the case of tag agreement on  \textit{$M_1$}. See Table \ref{box-now-table:tags} description of \textit{$M_1$}. Smaller counts here than might be expected from Fig.~\ref{fig:tagresults2} are explained by cases of one coder selecting sub-label \textit{Unfocused} after reaching $M_1$, while the other reached a code other than $M_1$ (cf.\ row $M_1$ of Table \ref{tab:heatmap_2tag}).}
    \label{fig:unclearaggreements}
    \vspace{-10pt}
\end{figure*}

\textit{Sub-label: Unfocused.}
While not affecting Q-nonagreements, Fig.~\ref{fig:unclearaggreements} shows that when coders agreed on tag $M_1$, they both chose \textit{Unfocused} in over 30\% of cases ($30+2$ of $106$ combined SS and SD cases), exactly one selected \textit{Unfocused} in $(13 + 3)/106 \approx 15$\% of cases, and neither selected \textit{Unfocused} in about 55\% of cases ($53 + 5$ of $106$). We expect these results reflect both the underlying dataset (likely a majority of items ended in $M_1$ for both coders because the items were unclear, i.e., not \textit{unambiguous language}, rather than not \textit{relatively focused}), and differing coder interpretations of the terminology in $Q_1$ including the meaning of \textit{relatively focused}; this is aside from it being somewhat unclear whether a coder is expected to select \textit{Unfocused} if they interpret an item to be both unclear \textit{and} not relatively focused. We comment on this further under `Recommendations'.

$Q_1$ also has the greatest difference in proportion of Q-nonagreements between cases SS and SD (35\% vs 17\%); here we also conjecture why this is, distinct from comparing $Q_1$ to other questions. The strongest idea is that when one coder tags an advice item with two tags (as occurs in SD), there are more opportunities for an agreement. To gain confidence in this intuition, we now briefly pursue the details. 
 
SS-type comparisons for $Q_1$ have 4 possible combinations for (C1)-(C2) answer pairs (N is \textit{no}, Y is \textit{yes}): 

\qquad Case A1: N-N 
\quad  A2: N-Y 
\quad  A3: Y-N
\quad  A4: Y-Y

\noindent A1 and A4 are Q-agreements. For the A2 and A3 Q-nonagreements, whichever coder answers \textit{no} (resulting in tag \textit{$M_1$}), the diverging question is $Q_1$ regardless of how many further questions are in the other coder's sequence. 
    
SD-type comparisons have 8 mathematically possible combinations for coders' $Q_1$ answers ($C1$)-($C2_1$, $C2_2$):

\qquad Case B1: N-YN$~~$ 
\quad  B2: N-YY$~~$ 
\quad  B3: Y-YN$~$ 
\quad  B4: Y-YY$~~$   

\qquad Case B5: N-NN* 
\quad  B6: N-NY$~~$ 
\quad  B7: Y-NN* 
\quad  B8: Y-NY \qquad *case cannot occur 

\noindent
Cases B5 and B7 cannot occur (NN from the two-tag coder would tag an advice item as \textit{$M_1$} twice). B6 (N-NY) and B8 (Y-NY) have the same outcome as B1 (N-YN) and B3 (Y-YN), as the order of C2's choices does not affect the outcome: Q-agreement occurs if either C2 answer matches C1's. In summary we have 6 cases for SD-type comparisons at $Q_1$: B1, B3, B4, B6, B8 are Q-agreements (on N, Y, Y, N, Y resp.), and B2 is a Q-nonagreement.

Thus a Q-nonagreement at $Q_1$ (for type SD) occurs only in one case of six (B2), and here the one-tag coder necessarily answered \textit{no}. The relatively constrained conditions of case B2, together with some expected variation in all answers (due to subjectivity and ambiguity, including in advice items), appears to explain the smaller proportion of Q-nonagreements at $Q_1$ for SD-type comparisons than for SS. As an aside, in case B2 (the SD-type comparison T-nonagreement case), regardless of how the two-tag coder answers questions beyond $Q_1$, the only node both coders visit is $Q_1$ (thus $Q_1$ is always the diverging question).

These constrained circumstances for SD-type nonagreements at $Q_1$ also provide a reason for a greater absolute number of SD-type Q-nonagreements in some later questions ($Q_3$, $Q_4$) than at $Q_1$ itself (right side of Fig.~\ref{fig:disagreementbar}A). 

\textbf{Question 3} (29\% and 23\% Q-nonagreements for type SS and SD, respectively).
$Q_3$ (Table \ref{nowTable:treeQuestions})  asks if an item is an outcome (vs.\ an action to take). Recall also this definition and $Q_3$'s annotation: 
    
\begin{displayquote}
    Outcome \cite{Barrera2021a, bellman2022a}: An \textit{outcome} is a result of some prior activity; in our context, often the end goal of advice.
\end{displayquote}
    
\begin{displayquote}
    $Q_3$ annotation \cite{Barrera2021a}: Is the advice a high-level outcome rather than some method (or meta-outcome) for how to achieve an outcome? E.g., \textit{data is secured in transit} would be an outcome because it is a desired goal or state, whereas \textit{data is encrypted in transit} is not because it explains (suggests) a method for achieving that outcome (in this case, encryption).\footnote{
        Advice phrased in past tense may at first glance be viewed as an outcome (e.g., \textit{data is encrypted in transit}). However, $Q_3$ and $Q_4$ of the coding tree specifically aim to distinguish outcomes from technical approaches (such as the use of encryption in this case). It is important, as noted by Barrera et al.~\cite{Barrera2021a}, to avoid mistakenly interpreting advice items as outcomes based purely on past-tense phrasing.
    } \textit{Encryption} may be considered a meta-outcome, as it is not meaningful to the end-user's ultimate goal of protected data.
\end{displayquote}
    
\noindent \textit{Meta-outcome} here suggests that encryption is not an end-goal itself, but rather a technical approach that provides a means to achieve a desired outcome (e.g., protected data).

If a coder finds this \textit{outcome} definition unclear, or cannot relate it to a given dataset item, they may be unsure how to answer $Q_3$. Note C2's tag set (Fig.\ref{fig:tagresults2}) contained \textit{T} (\textit{Outcome}) 2.4 times as often as C1 (235 vs. 97). Thus, often when C2 answered \textit{yes} to $Q_3$, C1's Q-nonagreement via \textit{no} yielded another tag from deeper in the tree.  

To explore this, we examined SS-type and SD-type tag nonagreements involving tag $T$, counting how often C2 used $T$ for an item while C1 assigned a lower tree tag (i.e., not $M_1$, $M_2$, or $T$; and $T$ was neither C1's first nor second tag).  Of $235$ times C2 used tag $T$, there were $142$ SS-type nonagreements, $106$ of which (75\%) involved tags other than $M_1$, $M_2$ from C1 (Table~\ref{tab:heatmap_2tag}, column $T$); and $27$ SD-type nonagreements,\footnote{
    Excluding $1$ for $M_1$, Table~\ref{tab:heatmapSD_3tag}'s column $T$ shows $44$ nonagreements (not $27$), as $2$ tag-vs-tag nonagrees are counted in the ($17$ of $27$) cases when C1 used two codes; $44 = (17 \times 2) + (27 - 17)$.
} $26$ of which (96\%) had C1 using a tag other than $M_1$, $M_2$. The remaining $235 - (142 + 27) = 66$ cases are accounted for by $58$ agreements on tag $T$ ($45$ SS-type, $13$ SD-type), and $8$ tag agreements where C2 selected $T$ as either first or second tag, but C2's other tag led to the agreement with C1 ($3$ of these involved DD-type comparisons, Table~\ref{tab:DDitems}).

This high proportion of occurrences ($106 + 26 = 132$ of $235$) of C1 assigning tags after $Q_4$ when C2 tags items as $T$ may suggest the description of \textit{outcome} is interpreted differently across coders, and that one or more of the definition of \textit{outcome}, the wording of $Q_3$ involving it, or $Q_3$'s extended annotation might be improved in materials provided to coders. 

The above definition of \textit{outcome} also relies on a mutual understanding of the term ``end goal'' shared by coders. If this differs, one coder may be more or less likely to answer \textit{yes} to $Q_3$. A brief description of \textit{desired end goal} is given in the annotation details \cite{Barrera2021a} using an example, but for more complex cases in the dataset, it may be unclear to coders whether a given item fits the definition. 
    
\textbf{Question 4} (14\% and 25\% Q-nonagreements for type SS and SD, respectively). $Q_4$ asks if an item suggests any of: a security technique, mechanism, software tool, or specific rule. This naturally relies on the definitions of these terms, which are not given in the question itself (but are given in part in annotations \cite{Barrera2021a} provided to coders, which include a few brief examples). An advice item might align poorly with these examples (if even they are consulted), resulting in coders implicitly relying on their own subjective definitions. This is discussed further under `Limitations (Clarity of terminology)'. 
    
Finally, we note that $Q_4$ is the main branch point where an advice item will either head toward actionable tags (plus \textit{$P_1$}), or toward more general advice such as principles. Q-nonagreements at $Q_4$ create a major divergence in coder sequences, in that resulting tags will differ even in terms of actionable vs.\ non-actionable (cf.\ Table~\ref{tab:disagreementsummary}).
    
\subsection{Low Numbers of Q-nonagreements}

As Fig.~\ref{fig:disagreementbar}A shows, $Q_2$, $Q_6$, $Q_7$, $Q_9$, and $Q_{10}$ have in total 6 Q-nonagreements across both comparison types (SS, SD). From these five, we now examine $Q_2$, $Q_6$, and $Q_7$ as the three that were most visited among these (per Fig.~\ref{fig:disagreementbar}B), and conjecture why they had so few or zero Q-nonagreements. While discussion is omitted, we note the number of times $Q_9$, $Q_{10}$  were respectively visited by both coders was relatively low, at ($59$, $18$) for SS-type and
($13$, $2$) for SD-type comparisons (again from Fig.~\ref{fig:disagreementbar}B).   
    
\textbf{Question 2} (1\% and 0\% Q-nonagreements for type SS and SD, respectively). $Q_2$ asks if the advice item is arguably helpful for security. For tag comparisons of type SS and SD, resp., $Q_2$ arose for 67\% (508/760, Fig.~\ref{fig:disagreementbar}B) and 88\% of items (206/234, Fig.~\ref{fig:disagreementbar}B), but was answered \textit{no} only 9 times. We first note the almost total absence of Q-nonagreements despite many opportunities. We next observe that almost all the advice was judged to have some possibility of improving security (we view this as more a reflection on the dataset, than the SAcoding method of analysis). Combined, this allows a conclusion that the dataset's advice items are, as expected, apropos to security (an easy hurdle), but provides less feedback on whether $Q_2$ would successfully filter out-of-scope advice in a mistargeted advice dataset, or would do so without unreasonably large numbers of Q-nonagreements. 

\textbf{Question 6} (0\% and 1\% Q-nonagreements for type SS and SD, respectively). $Q_6$ asks if the advice is viable to accomplish with reasonable resources. It was rarely answered \textit{no}, yielding \textit{$P_3$} in total 3 times (only one being counted as a Q-nonagreement at $Q_6$; cf.\ Tables \ref{tab:heatmap_2tag} and \ref{tab:heatmapSD_3tag}). This suggests nearly all actionable advice items in the 1013-item dataset are feasible to implement, per $Q_6$'s description of reasonable and Table \ref{box-now-table:tags}'s description of infeasible advice. However, we expect that the present dataset did not provide a strong test of $Q_6$ in general (similar to $Q_2$ above). A stronger test would be provided by datasets with more infeasible advice items. 
    
\textbf{Question 7} (0\% Q-nonagreements for both types). $Q_7$ asks if it is intended that the end-users of a product would be expected to carry out the advice item. As the target audience for the 1013-item dataset does not focus on end-users \cite{Barrera2021a}, it makes sense that few \textit{Specific Practice---End-User} (\textit{$P_6$}) tags would be used (\textit{yes} to $Q_7$), thus the absence of Q-nonagreements at $Q_7$ is not surprising. 
    
\subsection{Other Observations on Q-nonagreement Distributions}

As a rough measure, we expect the overall distribution for Q-nonagreements (Fig.~\ref{fig:disagreementbar}A) to be skewed towards lower-numbered questions, with the proportion of Q-nonagreements at subsequent nodes decreasing, as Q-nonagreements higher in the coding tree preclude later Q-nonagreements for a given advice item. For the proportion of Q-nonagreements within each question (Fig.~\ref{fig:disagreementbar}B) we do not expect this same pattern, as a Q-nonagreement higher in the tree has no impact on the proportion of Q-nonagreements within later questions (a reduced number of visits does not directly affect a proportion of nonagreements at a node). 

Reviewing Fig.~\ref{fig:disagreementbar}A and \ref{fig:disagreementbar}B, a few items draw our attention regarding the distribution of Q-nonagreements. Some questions have a small overall share of the Q-nonagreements  (Fig.~\ref{fig:disagreementbar}A), but the ratio of Q-nonagreements to total joint visits to that question is high (Fig.~\ref{fig:disagreementbar}B). One such case is $Q_8$, which we discuss next. A second such case is the now-removed question $Q_{11}$, as discussed under `Limitations'.

\textbf{Question 8} leads to two leaf nodes. The type SS results for overall Q-nonagreement proportion (Fig.~\ref{fig:disagreementbar}A, 8\%) and within-question Q-nonagreement proportion (Fig.~\ref{fig:disagreementbar}B, 34\%) indicate that when coders both reached $Q_8$, it was common for them to fail to agree. $Q_8$ asks about the level of security experience advice recipients are expected (or would need) to have. A coder with more security experience may answer differently than a less experienced coder; e.g., coders may apply personal security knowledge to distinguish groups (\textit{$P_4$/security expert} vs.\ \textit{$P_5$/IT specialist}). This may explain in part why a relatively large proportion of $Q_8$ answers were Q-nonagreements, as in our study the security experience of C1 (a senior PhD student in security at time of coding) and C2 (a professor with seven years of post-PhD security experience) was far from identical. 

Q-nonagreements at $Q_8$ cause concern, as we expect all IoT device manufacturers to employ IT specialists (\textit{$P_5$}) in order to develop devices, but fewer to have security experts (\textit{$P_4$}), especially in small and medium-sized companies. As such, which of the two audiences an advice item is actually designed or intended for is important, as items that target a security expert might not be helpful to (executable by) a non-expert. If the SAcoding method is able to identify advice items that do not match target recipients, then it offers a means to improve the targeting and thereby effectiveness of advice. 

\subsection{Comparing Actionable and Non-Actionable Tag Agreements}

To determine how well the coding tree can estimate the actionability of an advice dataset, we further calculated agreement within actionable tags, and within non-actionable tags (i.e., \textit{$P_3$}--\textit{$P_6$}, versus all others) for comparison types SS and SD. (Following previous work \cite{Barrera2021a}, we prioritize actionability as a desirable characteristic of security advice, while acknowledging that there is room for debate on this, depending on the particular advice item.) The method follows that for determining tag agreements and nonagreements as done earlier. This data is included in Table~\ref{tab:disagreementsummary} (column 4). 

For type SS, we consider the single tag of each of the two coders for a given advice item, and if both \textit{or} neither are an actionable tag, that counts as an actionability agreement; otherwise, a nonagreement. For type SS, coders agreed on 608 of 760 (80\% of) advice items (both actionable or both non-actionable). For type SD, we compare the single-tag coder's tag to each of the double-tag coder's tags. If in either pair, \{$C1$, $C2_1$\} or \{$C1$, $C2_2$\}, both tags are actionable or both are non-actionable, it counts as agreement; otherwise, a nonagreement. For type SD, coders agreed on 204 of 234 (87\% of) advice items regarding actionability. Unsurprisingly, this percentage is greater compared to type SS, because either of two pairs may yield an agreement. We view this as a limitation and give more weight here to the results from type SS. 

This analysis indicates that using the coding tree (and for this single large dataset), coders can distinguish an actionable practice (by the definition of Barrera et al.~\cite{Barrera2021a}) from a non-practice in over 80\% of cases---that is, actionable and non-actionable advice items are largely distinguishable. This does not necessarily, however, rule out tag nonagreements on \textit{which} practice code an item has (e.g., see $Q_8$ in Fig.~\ref{fig:disagreementbar}B). This nonetheless suggests the coding tree is useful for estimating how actionable an overall advice dataset is. 

\section{Deeper View of Tag Nonagreements via DCMS 13 Categories and Tag-vs-Tag Tables}


\begingroup

\definecolor{A}{RGB}{200, 200, 200}
\definecolor{B}{RGB}{190, 190, 190}
\definecolor{C}{RGB}{180, 180, 180}
\definecolor{D}{RGB}{170, 170, 170}
\definecolor{E}{RGB}{160, 160, 160}
\definecolor{F}{RGB}{150, 150, 150}
\definecolor{G}{RGB}{140, 140, 140}
\definecolor{H}{RGB}{130, 130, 130}
\definecolor{I}{RGB}{120, 120, 120}
\definecolor{J}{RGB}{110, 110, 110}
\definecolor{K}{RGB}{100, 100, 100}
\definecolor{L}{RGB}{90, 90, 90}
\definecolor{M}{RGB}{80, 80, 80}

\setlength{\tabcolsep}{2pt}
\renewcommand{\arraystretch}{1}

\begin{table}[!t]
    {
        \centering
        \caption{Distribution of codes assigned to advice items for coders \textit{C1} and \textit{C2}, grouped into 13 DCMS topics or categories (using the mapping document \cite{DCMS1}). Cell values give percentage of each category's items tagged with column 1 codes. Column sums for C1 or C2 may exceed 100\% as coders could optionally select a second code for an item. SP is short for Specific Practice. For context and reader convenience, titles of the 13 categories UK-i are given in Table~\ref{tab:guidelines}. 
        \vspace{-1pt}}
        \label{tab:c1vsc2}
        \small
        \begin{tabular}{llrrrrrrrrrrrrrr}
			& & \multicolumn{13}{c}{UK Guideline category (and number of items in category, from 1013 total)} \\ \cline{3-15}
			Code & C & UK-1 & UK-2 & UK-3 & UK-4 & UK-5 & UK-6 & UK-7 & UK-8 & UK-9 & UK-10 & UK-11 & UK-12 & UK-13 \\
			\cline{3-15}
			   & & 81   & 63   & 145  & 84   & 165  & 161  & 65   & 98   & 39   & 51    & 22    & 18    & 21    \\
			\toprule
			\multirow{2}{*}{$M_1$. Not Useful} 
			& C1 & \cellcolor{D}17\% & \cellcolor{G}33\% & \cellcolor{F}30\% & \cellcolor{D}20\% & \cellcolor{E}24\% & \cellcolor{D}17\% & \cellcolor{F}28\% & \cellcolor{E}21\% & \cellcolor{I}\color{white}{41}\% & \cellcolor{J}\color{white}{49}\% & \cellcolor{B}9\%  & \cellcolor{E}22\% & \cellcolor{B}10\%  \\
			& C2 & \cellcolor{A}5\%  & \cellcolor{C}14\% & \cellcolor{C}13\% & \cellcolor{C}11\% & \cellcolor{D}16\% & \cellcolor{C}12\% & \cellcolor{C}12\% & \cellcolor{C}14\% & \cellcolor{E}21\% & \cellcolor{F}29\% & \cellcolor{A}5\%  & \cellcolor{C}11\% & \cellcolor{B}10\%  \\
			\cline{1-15}
			\multirow{2}{*}{$M_2$. Beyond Scope} 
			& C1 & 0\%  & 0\%  & 0\%  & 0\%  & 0\%  & 0\%  & 0\%  & 0\%  & 0\%  & 0\%  & 0\%  & 0\%  & 0\%  \\
			& C2 & 0\%  & \cellcolor{A}5\%  & \cellcolor{A}1\%  & 0\%  & \cellcolor{A}1\%  & 0\%  & 0\%  & \cellcolor{A}2\%  & 0\%  & 0\%  & 0\%  & \cellcolor{B}6\%  & 0\%  \\
			\cline{1-15}
			\multirow{2}{*}{$N$. Security Principle} 
			& C1 & 0\%  	& \cellcolor{A}2\%  & \cellcolor{A}2\%  	& \cellcolor{A}1\%  & \cellcolor{A}1\%	& \cellcolor{B}10\%  & \cellcolor{A}5\%  & \cellcolor{B}6\%  & \cellcolor{B}8\%  & 0\%  & 0\%  & \cellcolor{B}6\%	& 0\%  \\
			& C2 & 0\%  	& 0\%  				& \cellcolor{A}1\%		& \cellcolor{A}2\%	& \cellcolor{A}5\%  & \cellcolor{D}16\% & \cellcolor{A}5\%  & \cellcolor{A}3\%  & \cellcolor{B}8\%  & 0\%  & 0\%  & \cellcolor{C}11\%	& \cellcolor{C}15\%  \\
			\cline{1-15}
            \multirow{2}{*}{\shortstack[l]{$P_1$. Incompletely\\  $~~~~~$Specified Practice}} 
			& C1 & \cellcolor{C}14\% & \cellcolor{C}11\% & \cellcolor{E}23\% & \cellcolor{G}32\% & \cellcolor{G}33\% & \cellcolor{F}27\% & \cellcolor{E}25\% & \cellcolor{D}18\% & \cellcolor{G}31\% & \cellcolor{G}33\% & \cellcolor{G}32\% & \cellcolor{D}17\% & \cellcolor{G}33\% \\
			& C2 & \cellcolor{D}20\% & 0\%  & \cellcolor{C}13\% & \cellcolor{D}19\% & \cellcolor{E}22\% & \cellcolor{D}16\% & \cellcolor{E}22\% & \cellcolor{C}11\% & \cellcolor{B}10\% & \cellcolor{D}20\% & \cellcolor{B}9\%  & 0\%  & \cellcolor{B}10\%  \\ 
			\cline{1-15}
			\multirow{2}{*}{\shortstack[l]{$P_2$. General Practice/\\ $~~~~~$Policy}}
			& C1 & \cellcolor{A}1\%  & \cellcolor{I}\color{white}{43}\% & \cellcolor{D}17\% &0\%  & \cellcolor{A}1\%  & \cellcolor{A}2\%  & 0\%  & \cellcolor{G}35\% & \cellcolor{A}5\%  & \cellcolor{A}4\%  & \cellcolor{D}18\% & \cellcolor{F}28\% & 0\%  \\
			& C2 & \cellcolor{A}1\%  & \cellcolor{L}\color{white}{57}\% & \cellcolor{C}15\% & \cellcolor{A}2\%  & \cellcolor{A}3\%  & \cellcolor{A}3\%  & \cellcolor{A}3\%  & \cellcolor{F}27\% & \cellcolor{A}5\%  & \cellcolor{C}12\% & \cellcolor{C}14\% & \cellcolor{D}17\% & \cellcolor{A}5\%  \\ 
			\cline{1-15}
			\multirow{2}{*}{$P_3$. Infeasible Practice} 
			& C1 & 0\%  & 0\%  & 0\%  & 0\%  & 0\%  & 0\%  & 0\%  & 0\%  & 0\%  & 0\%  & 0\%  & 0\%  & 0\%  \\
			& C2 & 0\%  & 0\%  & 0\%  & 0\%  & 0\%  & \cellcolor{A}1\%  & 0\%  & \cellcolor{A}1\%  & 0\%  & 0\%  & 0\%  & 0\%  & 0\%  \\ 
			\cline{1-15}
			\multirow{2}{*}{$P_4$. (SP) Security Expert} 
			& C1 & \cellcolor{D}17\% & \cellcolor{A}2\%  & \cellcolor{C}12\% & \cellcolor{H}37\% & \cellcolor{F}27\% & \cellcolor{C}15\% & \cellcolor{E}23\% & \cellcolor{A}1\%  & \cellcolor{A}5\%  & \cellcolor{A}4\%  & \cellcolor{A}5\%  & \cellcolor{B}6\%  & \cellcolor{F}29\% \\
			& C2 & \cellcolor{G}31\% & \cellcolor{A}3\%  & \cellcolor{C}14\% & \cellcolor{I}\color{white}{44}\% & \cellcolor{E}24\% & \cellcolor{C}14\% & \cellcolor{E}25\% & \cellcolor{A}4\%  & \cellcolor{B}8\%  & \cellcolor{B}6\%  & \cellcolor{A}5\%  & 0\%  & \cellcolor{G}33\% \\ 
			\cline{1-15}
			\multirow{2}{*}{$P_5$. (SP) IT Specialist} 
			& C1 & \cellcolor{I}\color{white}{43}\% & \cellcolor{B}8\%  & \cellcolor{A}5\%  & \cellcolor{B}7\%  & \cellcolor{B}7\%  & \cellcolor{D}17\% & \cellcolor{A}5\%  & \cellcolor{A}5\%  & \cellcolor{A}3\%  & 0\%  & \cellcolor{B}9\%  & \cellcolor{C}11\% & \cellcolor{F}29\% \\ 
			& C2 & \cellcolor{G}32\% & \cellcolor{B}6\%  & \cellcolor{B}8\%  & \cellcolor{B}6\%  & \cellcolor{B}7\%  & \cellcolor{C}15\% & \cellcolor{B}9\%  & \cellcolor{B}8\%  & \cellcolor{E}23\% & \cellcolor{B}6\%  & \cellcolor{F}27\% & \cellcolor{C}11\% & \cellcolor{D}19\% \\ 
			\cline{1-15}
			\multirow{2}{*}{$P_6$. (SP) End-User} 
			& C1 & \cellcolor{A}1\%  & 0\%  & 0\%  & 0\%  & 0\%  & 0\%  & 0\%  & \cellcolor{A}1\%  & 0\%  & 0\%  & 0\%  & 0\%  & 0\%  \\
			& C2 & 0\%  & 0\%  & \cellcolor{A}1\%  & 0\%  & 0\%  & 0\%  & 0\%  & \cellcolor{A}1\%  & 0\%  & 0\%  & 0\%  & 0\%  & 0\%  \\
			\cline{1-15}
			\multirow{2}{*}{$T$. Desired Outcome} 
			& C1 & \cellcolor{B}6\%  & \cellcolor{A}2\%  & \cellcolor{B}10\% & \cellcolor{A}2\%  & \cellcolor{B}7\%  & \cellcolor{C}12\% & \cellcolor{C}15\% & \cellcolor{C}12\% & \cellcolor{B}8\%  & \cellcolor{B}10\%  & \cellcolor{F}27\% & \cellcolor{B}6\%  & 0\%  \\
			& C2 & \cellcolor{B}9\%  & \cellcolor{C}14\% & \cellcolor{G}35\% & \cellcolor{C}14\% & \cellcolor{D}17\% & \cellcolor{E}22\% & \cellcolor{E}25\% & \cellcolor{F}28\% & \cellcolor{F}26\% & \cellcolor{F}28\% & \cellcolor{G}32\% & \cellcolor{I}\color{white}{44}\% & \cellcolor{B}10\%  \\
			\cline{1-15}
			\multirow{2}{*}{$T'$. Desired Outcome} 
			& C1 & 0\%  & 0\%  & 0\%  & 0\%  & 0\%  & 0\%  & 0\%  & 0\%  & 0\%  & 0\%  & 0\%  & \cellcolor{B}6\%  & 0\%  \\
			& C2 & \cellcolor{A}3\%  & 0\%  & \cellcolor{A}1\%  & \cellcolor{A}1\%  & \cellcolor{A}4\%  & \cellcolor{A}1\%  & 0\%  & \cellcolor{A}1\%  & 0\%  & 0\%  & \cellcolor{B}9\%  & 0\%  & 0\%  \\
			\bottomrule
            \multirow{2}{*}{\% tagged as actionable} 
            & C1 & \cellcolor{M}\color{white}{62}\% & \cellcolor{B}10\%  & \cellcolor{D}17\% & \cellcolor{I}\color{white}{44}\% & \cellcolor{G}34\% & \cellcolor{G}32\% & \cellcolor{F}28\% & \cellcolor{B}7\%  & \cellcolor{B}8\%  & \cellcolor{A}4\%  & \cellcolor{C}14\% & \cellcolor{D}17\% & \cellcolor{L}\color{white}{57}\% \\
            & C2 & \cellcolor{M}\color{white}{63}\% & \cellcolor{B}10\%  & \cellcolor{F}22\% & \cellcolor{J}\color{white}{50}\% & \cellcolor{G}32\% & \cellcolor{F}29\% & \cellcolor{G}34\% & \cellcolor{C}14\% & \cellcolor{G}31\% & \cellcolor{C}12\% & \cellcolor{G}32\% & \cellcolor{C}11\% & \cellcolor{K}\color{white}{52}\% \\
            \midrule
            C1--C2 actionability delta 
            & & \cellcolor{A}1\% & 0\% & \cellcolor{A}5\% & \cellcolor{B}6\% & \cellcolor{A}2\% & \cellcolor{A}3\% & \cellcolor{B}6\% & \cellcolor{B}7\% & \cellcolor{E}23\% & \cellcolor{B}8\% & \cellcolor{D}18\% & \cellcolor{B}6\% & \cellcolor{A}5\% \\
            \bottomrule
		\end{tabular}
	}
\end{table}

\endgroup

\begin{table}[htb!]
    \centering
    \caption{Titles of 13 UK DCMS guidelines of Table \ref{tab:c1vsc2}. Full guideline descriptions can be found in the original document~\cite{DCMS2}.} 
    \label{tab:guidelines}
    \small{
        \begin{tabular}{ll}
            Guideline & Title \\  \midrule
            UK-1 & \textit{No default passwords} \\
            UK-2 & \textit{Implement a vulnerability disclosure policy} \\
            UK-3 & \textit{Keep software updated}  \\
            UK-4 & \textit{Securely store credentials and security-sensitive data} \\
            UK-5 & \textit{Communicate securely}  \\
            UK-6 & \textit{Minimise exposed attack surfaces} \\
            UK-7 & \textit{Ensure software integrity}  \\
            UK-8 & \textit{Ensure that personal data is protected} \\
            UK-9 & \textit{Make systems resilient to outages} \\
            UK-10 & \textit{Monitor system telemetry data} \\
            UK-11 & \textit{Make it easy for consumers to delete personal data}  \\
            UK-12 & \textit{Make installation and maintenance of devices easy} \\
            UK-13 & \textit{Validate input data}   \\
        \end{tabular}
    }
\end{table}

As visible from Fig.~\ref{fig:tagresults2} at the beginning of our analysis, for some tags, there are large differences in the frequency of tag occurrence between coders. Table~\ref{tab:c1vsc2} gives a different view of the differences in tag assignments, showing what proportion of items in each of 13 categories was assigned each tag by each coder; this partitioning into 13 bins sharply signals that similar proportions of items being given a tag by each coder does not imply that each coder gave the tag to the same individual items. The 13 categories are from a UK government effort \cite{DCMS2} which we refer to as the DCMS 13 guidelines, with titles as given in   Table~\ref{tab:guidelines}. Each of the 1013 advice items under discussion are mapped to one of these 13 categories by a ``mapping'' document \cite{DCMS1} used in conjunction with the DCMS 13 guidelines. Table~\ref{tab:c1vsc2} reveals the codes most commonly assigned to each advice category (topic) by C1 and C2; its heatmap shading aims to aid comparing the coders' code distributions. 

For example, among the 1013 advice items that fall in the (table column) category of  \textit{UK-10}, C1 tagged 49\% of the items as \textit{$M_1$}, while C2 assigned this tag to 29\% of items. Proportions of other tags within this same column also differ considerably between C1 and C2. For example, in the row for tag \textit{T}, the respective values are 10\% and 28\%. Here, \textit{$M_1$} arose more frequently for C1 (than C2), while \textit{T} resulted more frequently for C2 (than C1).

Although we do not have evidence of any causal relationship in this pairing of \textit{T} and \textit{$M_1$}, the possibility of such relationships between pairs of codes, and large cross-coder differences visible from Table~\ref{tab:c1vsc2}, motivate deeper analysis. An example of a concrete question that we would like to answer is the following. Consider all items that C1 tagged with a particular code, say $P_1$, and that result in tag nonagreements; how are C2's (nonagreeing) tags for this subset of items distributed over the other tag values, and are there patterns worth exploring? In addition, Table~\ref{tab:disagreementsummary} shows tag agreements for (only) 46\% of tags ($315 + 130 + 17 = 462$ of 1013 items); this is lower than we expected from earlier-reported testset results between three coders \cite{Barrera2021a} (summarized herein at the end of Related Work), albeit those testset results counted near-matches as matches through the so-called $\pm 1$ rule, which is not used herein. Nonetheless, these issues motivate deeper exploration to better understand tag nonagreements. 

To pursue this, we introduce \textit{tag-vs-tag nonagreement tables} (Tables \ref{tab:heatmap_2tag} and \ref{tab:heatmapSD_3tag}). Table \ref{tab:heatmap_2tag} shows the distribution of tags for SS-type tag nonagreements. For example, its row $T$ tells us that for the items tagged $T$ by C1, there were 26 tag nonagreements in total, and these nonagreements were distributed across 6 different tags (identified by respective column headers), which C2 selected with frequency 4, 6, 5, 2, 4, and 5.

\begin{table}[b]
\centering
\caption{Tag-vs-Tag table for SS-type tag nonagreements. For a fixed row, e.g., $P_1$ (implying advice items C1 single-tagged as $P_1$ but that C2 did not tag as $P_1$), the columns show the distribution of C2's (nonagreeing, single) tags, showing how many of these items C2 tagged with the code of that column's heading. For a fixed column, reverse the roles of C1 and C2 in this description. Sum of all counts is 445 (matching the number of SS-type nonagreements in Fig.~\ref{fig:disagreementbar}). Dash (--) represents 0.}
    \label{tab:heatmap_2tag}
    \small
    \begin{tabular}{ll@{}rrrrrrrrrrrrr}
    &     & \multicolumn{11}{c}{Coder 2 codes} & \\
    \cline{3-13}
    &     & $M_1$ & $M_2$ & $N$  & $P_1$ & $P_2$ & $P_3$ & $P_4$ & $P_5$ & $P_6$ & $T$   & $T'$ & Sum \\ \cline{3-13}
    \parbox[t]{2mm}{\multirow{11}{*}{\rotatebox[origin=c]{90}{Coder 1 codes}}}
    &$M_1$   & *  & 3  & 5  & 21 & 19 & -- & 33 & 12 & -- & 36  & 3  & 132 \\
    &$M_2$   & -- & *  & -- & -- & -- & -- & -- & -- & -- & --  & -- & 0   \\
    &$N$    & 2  & -- & *  & 2  & -- & -- & -- & -- & -- & 7   & -- & 11  \\
    &$P_1$   & 11 & -- & 12 & *  & 16 &1  & 24 & 9  & -- & 61  & 4  & 138 \\
    &$P_2$   & 6  & 4  & -- & 3  & *  & -- & 2  & 11 & -- & 21  & -- & 47  \\
    &$P_3$   & -- & -- & -- & -- & -- & *  & -- & -- & -- & --  & -- & 0   \\
    &$P_4$   & 1  & -- & 5  & 16 & 1  & -- & *  & 11 & -- & 13  & -- & 47  \\
    &$P_5$   & -- & -- & 2  & 7  & 6  & -- & 25 & *  & -- & 3   & -- & 43  \\
    &$P_6$   & -- & -- & -- & -- & -- & -- & -- & -- & *  & --  & -- & 0   \\
    &$T$    & 4  & -- & -- & 6  & 5  & -- & 2  & 4  & -- & *   & 5  & 26  \\
    &$T'$ & -- & -- & -- & -- & -- & -- & -- & -- & -- & 1   & *  & 1   \\
    \cline{3-13}
    &Sum & 24 & 7  & 24 & 55 & 47 & 1  & 86 & 47 & 0  & 142 & 12 &     \\
    
    \end{tabular}
    \vspace{-5pt}
\end{table}

These tables allow an understanding of aspects from Fig.~\ref{fig:tagresults2} earlier in the paper that drew our attention, four of which we note here: the relatively large cross-coder differences in proportion of items receiving tags $T$, $P_1$, and $M_1$, plus C2 using $N$ almost twice as often as C1. While each of these gaps suggest approximate lower bounds on the numbers of resulting tag nonagreements, how much higher the actual numbers are is not revealed---for example, C1 assigned $T$ to far fewer items than C2, yet some of these may be distinct from items tagged $T$ by C2 (i.e., coders may have applied $T$ to different items). Beyond the proportions shown by Fig.~\ref{fig:tagresults2}, tag-vs-tag tables reveal the actual numbers of tag nonagreements, and the relationships between each pair of tags for such nonagreements. 

Scanning the largest entries in Table \ref{tab:heatmap_2tag}, we see that relatively large numbers of SS tag nonagreements occur in particular when C1 tags items as $M_1$ or $P_1$ (rows), and when C2 tags items as $T$ (column). We can also track `where the nonagreeing tags go', including for each of the four cases noted in the preceding paragraph.

Table \ref{tab:heatmap_2tag} reveals a fifth case where a relatively large number of tag nonagreements occur: sets of items that C2 tags as $P_4$ (column). This case was not apparent from Fig.~\ref{fig:tagresults2}, as C1 and C2 tagged similar proportions of items with $P_4$; however, this detailed table reveals that many $P_4$ tags were on different items for C1 than for C2.

In Table \ref{tab:heatmap_2tag} overall, the largest non-sum entry is 61, where row $P_1$ intersects column $T$. This means the greatest number of  tag nonagreements occurred for items tagged $P_1$ by C1, while $T$ by C2; this is perhaps explained by, for some advice items, an overlap between an \textit{Outcome} (missing details on how to reach it) and an \textit{Incompletely Specified Practice}, depending on item wording and coder interpretation of terms. Also note: the four entries in the corners of the sub-table bounded by rows $M_1$ and $P_1$ (the first 4 rows), and columns $P_4$ and $T$, alone sum to $33 + 36 + 24 + 61 = 154$ of $445$ total SS-type tag nonagreements (35\%); and C1's rows $M_1$ and $P_1$ in total account for $132 + 138 = 270$ of $445$ SS nonagreements (61\%). Turning to C2, column's $P_4$ and $T$ together account for $86 + 142 = 228$ of $445$ SS nonagreements (51\%). Thus the table reveals that a small number of conditions account for substantial portions of all tag nonagreements.

Another outlier in Table \ref{tab:heatmap_2tag} is value 25 at row $P_5$ (\textit{IT Specialist}), column $P_4$ (\textit{Security Expert}). This might be explained by different interpretations across coders of the capabilities or knowledge of these closely-related expert classes. On a similar note, value 21 (row $P_2$, col $T$) might be characterized as one coder recognizing as a \textit{general approach} what another considers to be an \textit{outcome}; note that both codes imply a lack of stepwise details, and an item suggesting a general approach might also hint at a desired outcome, depending on advice item wording. 

As a further example from Table \ref{tab:heatmap_2tag}, to explore the earlier-noted result (visible from Fig.~\ref{fig:tagresults2}) that C2 chooses $N$ twice as often as C1, we ask: When this results in a tag nonagreement, what did C1 choose most frequently? To answer, from column $N$ we find the greatest number of nonagreements is row $P_1$ (12 times, 50\% of this column's nonagreements), where C1 tagged items as  \textit{Incompletely Specified Practice}. In retrospect, this is perhaps not surprising: when C1 interprets an item as this, C2 often coded the same item as a \textit{Principle}. Again notice for items matching both these codes, stepwise details would typically be missing. Note also that in earlier work, these codes ($P_1$, $N$) were placed adjacent on a relationship continuum (Fig.3 in \cite{Barrera2021a}) ranging from focus on end-result to focus on mechanism details---suggesting that despite T-nonagreements here, the two coders are conceptually close. Finally, note that from Table \ref{tab:heatmap_2tag} we can observe: 24 nonagreements occurred when C2 chose $N$, and 11 nonagreements occurred when C1 chose $N$, for a total of exactly 35 SS-type nonagreements; while from Table \ref{tab:heatmapSD_3tag} (below), there are a further $5 + 14 = 19$ SD-type nonagreements related to one or the other using $N$, for a sum of $35 + 19 = 54$. In contrast, what can be deduced from the bars for $N$ in Fig.~\ref{fig:tagresults2} is that the number of nonagreements related to coders choosing code $N$ has approximate lower and upper bounds of $59 - 35 = 24$ (C2's count minus C1's) and $59 + 35 = 94$ (C2's count plus C1's).

\begin{table}[t!]
\centering
\caption{Tag-vs-Tag table for SD-type tag nonagreements. See Table \ref{tab:heatmap_2tag} for explanation. Sum of all counts is 208 \textit{tag-vs-tag nonagreements}, twice the 104 tag nonagreements implied by Table~\ref{tab:disagreementsummary}, because for items with SD-type nonagreements, each tag from the 2-tag coder yields a tag-vs-tag nonagreement. Note from Table~\ref{tab:disagreementsummary}: $234$ (SD-type comparison items) $- 130 = 104$.}
\label{tab:heatmapSD_3tag}
\small
\begin{tabular}{llrrrrrrrrrrrrr}
    &     & \multicolumn{11}{c}{Coder 2 codes} & \\
    \cline{3-13}
    &     & $M_1$ & $M_2$ & $N$  & $P_1$ & $P_2$ & $P_3$ & $P_4$ & $P_5$ & $P_6$ & $T$   & $T'$ & Sum \\ \cline{3-13}
    \parbox[t]{2mm}{\multirow{11}{*}{\rotatebox[origin=c]{90}{Coder 1 codes}}}
    &$M_1$  & *  & -- & -- & 9  & 2  & -- & 15 & 4  & -- & 1  & -- & 31  \\
    &$M_2$  & -- & *  & -- & -- & -- & -- & -- & -- & -- & -- & -- & 0   \\
    &$N$   & -- & -- & *  & 2  & 1  & -- & 1  & -- & -- & 1  & -- & 5   \\
    &$P_1$  & 3  & -- & 6  & *  & 10 & 2  & 7  & 17 & 1  & 17 & 5  & 68  \\
    &$P_2$  & 2  & 2  & -- & 2  & *  & -- & -- & 2  & -- & 4  & -- & 12  \\
    &$P_3$  & -- & -- & -- & -- & -- & *  & -- & -- & -- & -- & -- & 0   \\
    &$P_4$  & 3  & -- & 4  & 7  & 8  & 1  & *  & 11 & -- & 12 & 2  & 48  \\
    &$P_5$  & 1  & -- & 3  & 7  & 3  & -- & 5  & *  & -- & 10 & -- & 29  \\
    &$P_6$  & -- & -- & -- & -- & -- & -- & 1  & -- & *  & -- & -- & 1   \\
    &$T$   & 1  & -- & 1  & 3  & -- & -- & 4  & 3  & -- & *  & 2  & 14  \\
    &$T'$  & -- & -- & -- & -- & -- & -- & -- & -- & -- & -- & *  & 0   \\
    \cline{3-13}
    &Sum & 10 & 2  & 14 & 30 & 24 & 3  & 33 & 37 & 1  & 45 & 9  &     \\
    \end{tabular}
    \vspace{-5pt}
\end{table}

Table \ref{tab:heatmapSD_3tag} is for SD-type tag nonagreements; given the lengthy discussion of Table \ref{tab:heatmap_2tag}, here we are intentionally brief. Table \ref{tab:heatmapSD_3tag} provides somewhat less information, there being fewer nonagreements (104) of this type. The total of the counts in this table is $2 \times 104 = 208$, because for each nonagreement, there are two tags from one coder that fail to agree with the single tag from the other (as noted also in the table caption). Here the entries that draw our greatest attention are the sums (68) for row $P_1$, and (45) for column $T$. From data given in the table itself, individual cases of interest can be explored independently, as for Table \ref{tab:heatmap_2tag} above.

\section{Utility, Limitations, and Recommendations}

We begin this section with a summary of areas where we believe the SAcoding method has been shown to be useful. We then continue on to discuss limitations, combined with  recommendations.

\subsection{Utility of the SAcoding Method}

We now sum up our view of the utility of the SAcoding method, and benefits it offers to the security community.

\textbf{Categorizing security advice datasets.} 
Applying the method to analyze the 1013-item advice dataset (believed to broadly represents IoT security advice) allowed a characterization of the dataset. This included, for the advice items offered, the comparative number of items in designated categories (outcomes, principles, practices, policies, and those unclassifiable). The tagging by two separate coders resulted in similar proportions of tags for many, but far from all, tag categories; however our confidence in this result is weak, as more detailed analysis revealed that the coders often did not assign the same tag to a given individual item. We thus find it premature to make a broad claim that the method is useful for identifying advice that belongs to the high-level categories represented by the tags, or for accurate estimates of the overall proportion of advice that falls into each category. 
  
\textbf{Identifying non-actionability of advice.} 
Applying the method to the 1013-item advice dataset identified a significant proportion of current IoT security advice to be non-actionable. The method thus appears to be useful to identify non-actionable advice in a dataset. While its use in the present paper is scoped to this representative IoT advice dataset, we believe that both advice givers and advice targets (recipients) in other security areas may also benefit from the method. (We note that while the coding interface allowed ticking a checkbox to denote items as IoT-specific,
neither C1 nor C2 selected this for any of 1013 items.)
        
\textbf{Measuring and characterizing actionability.} 
The SAcoding method helps measure not only what proportion of advice in a dataset is actionable, but in characterizing \textit{why} an individual advice item is (or is not) actionable. As the tree directs coders toward tags that characterize a given advice item, the questions at each node along the path contribute to a characterization or description of the assigned tag. For tags that are actionable (all tags after $Q_6$), the answers to questions leading to those tags characterize each practice tag and what makes them actionable. These characteristics include whether the advice provides a technique or mechanism to use ($Q_4$), how technically detailed the advice is ($Q_5$), and which target audience the advice is scoped to ($Q_7$ and $Q_8$). In the present paper these properties are used to characterize actionable advice in one large dataset; in other work \cite{bellman2022a}, this is used to compare actionability of distinct, smaller datasets.

\textbf{Cross-checking target recipients and target codes.}
The method would appear to be useful for advice givers (dataset creators) to cross-check the following, for advice datasets under creation:
\begin{enumerate}
    \item that intended target audiences match coding results regarding targets implied by codes $P_4$--$P_6$; 
    \item that advice item wording delivers the higher-level specific types of advice intended by advice givers, with respect to advice categories of actionable practices, principles, approaches/policies, or outcomes; and
    \item that items falling into code
    $M_1$ (unclear or unfocused) are either clarified or split into finer-grained items, as appropriate; those falling into
    $M_2$ (not security related) are removed; and those falling into $P_1$ (incompletely specified) and $P_3$ (infeasible) are appropriately revised. These codes are the dark-shaded items in Fig.~\ref{fig:flowchart2}.
\end{enumerate}         

\subsection{Limitations and Recommendations}

Based on our analysis, here we summarize our view of limitations both of the SAcoding method and our analysis of it. We also give recommendations on how it might be modified, and thoughts on further work.

\textbf{Findings based on 1013-item dataset.} 
While we expect the coding tree to be useful on security advice datasets beyond IoT (as noted above)---its questions were designed to be generic and applicable beyond IoT security advice---our findings herein are largely based on one 1013-item dataset \cite{BellmanDataset}. Lower-numbered questions in the coding tree are broad filters to determine if advice is suitable for analysis (language-wise, and based on focus); later questions are finer-grained (e.g., $Q_8$ differentiates practices intended for security experts from those for IT specialists), and $Q_6$ asks whether a practice is viable with reasonable resources (splitting off practices considered infeasible). However, to date, the method has not been tested beyond IoT security advice datasets.  

\textbf{Number of coders.} 
For our cross-check of coding, a second coder (C2) coded the full dataset. The degree of coder agreement on tag distribution (i.e., proportions) appeared promising, but confidence weakened upon more detailed tag-vs-tag analysis. Agreement on advice actionability was more promising (80\% and 87\% agreement on individual items being assigned actionable tags or not, for SS and SD cases, resp.). Given the mixed results, and the limitation of a single pair of coders (and one fixed dataset, albeit large), further analysis with additional coders would provide stronger confidence in the method if this delivered comparable or better results; it could also serve to test out ideas for improving the SAcoding method itself. A practical challenge for large datasets such as ours is the substantial human time cost per coder (for each advice item, reading it and answering up to 10 questions). One approach to reduce the time burden on individual coders is to use larger numbers of coders each tagging a randomly-drawn subset or partition from a full dataset; however, this may introduce unanticipated side effects, or preclude useful analysis of tagging patterns, e.g., related to coder demographics or experience levels.

Investigating agreement between three or more coders is beyond the scope of the present paper; we encourage further investigation of coder agreement and approaches to overcome the challenges mentioned here.

\textbf{$T$--$T'$ agreement.} In our analysis, tags $T$ and $T'$ are considered a nonagreement; this allows tracking their distinct coding paths (origins). Such nonagreements occurred $6+2$ times for SS+SD cases, per Tables~\ref{tab:heatmap_2tag} and \ref{tab:heatmapSD_3tag}. As both tags have the same word description, one could argue for treating them in analysis as a single code. The pros and cons of this design/analysis choice, and its impact on results, could be explored further.

\textbf{Extracting sub-topics before tagging.} 
As mentioned earlier, a notable number of items (14\% and 5\% for C1 and C2, resp.) were tagged with sub-label \textit{Unfocused} (as an option of \textit{$M_1$}), implying the coders found multiple sub-topics. Prior to tagging an item with multiple sub-topics, if those sub-topics were extracted for individual tagging (vs.\ as a composite item), each  might well receive a tag other than \textit{$M_1$}. This of course would change both the number of items in the dataset, and the results. This idea, which can allow more meaningful feedback to those creating advice datasets, is pursued in other work \cite{bellman2022a}; this also overlaps the next point. 

\textbf{Splitting $Q_1$ (and $M_1$).}
A recommended coding tree refinement is to split $Q_1$ into two  questions. A new first question $Q_{1A}$ would ask whether the advice item is unfocused (e.g., multiple items on different topics, or too wide a variety of aspects of the same topic to be meaningfully assigned a single tag---in this case the item might be split in a pre-processing step to enable meaningful tagging); a \textit{yes} here  would result in a new code $M_{1A}$, elevating the current sub-label \textit{Unfocused} to a (regular) code. A \textit{no} would proceed to $Q_{1B}$, asking whether the advice item is unclear; there a \textit{yes} results in a new assigned code $M_{1B}$ for \textit{Unclear} (grammar issues, language ambiguity). We believe this would address ambiguity that arose in trying to understand the meaning of coder choices in our earlier interpretation of Question 1 and Fig.~\ref{fig:unclearaggreements}. 

\textbf{Combined $N$ tag for design principles.} 
The coding tree of Fig.\ref{fig:flowchart2} is minorly simplified from prior work~\cite{Barrera2021a}, which included a $Q_{11}$ (asking whether the advice item relates to the design phase of the product lifecycle), and thus had separate tags $N1$ (\textit{security principle}) and $N2$ (\textit{security design principle}, as a subset). In fact, coder C2 used the original method with old $Q_{11}$ and these separate tags, though our analysis to this point combined these into the single tag $N$ (principle), with items coded $N1$ or $N2$ combined and counted as $N$. Eliminating $Q_{11}$ and using $N$ simplifies the coding process minorly (one fewer question), without meaningful loss of information---as we now explain, from a brief separate analysis (next) retaining $Q_{11}$ and separate codes $N1$, $N2$.

Old $Q_{11}$ was reached 94 times in total counting visits by C1 and C2 separately---relatively infrequent, but not negligible. The main point however is that we found that $Q_{11}$ and separate codes $N1$ and $N2$ provided little useful information. \textbf{$Q_{11}$} was reached by \textit{both} coders relatively infrequently ($18+2$ times for SS+SD cases), but when it was, Q-nonagreement occurred in 39\% of SS-type comparison instances ($7$ of $18$), and $0$ of $2$ SD-type instances; this comprised 2\% of Q-nonagreements over all questions. As $Q_{11}$'s answer is based on coders' view of where in the device lifecycle \cite{Barrera2021a} the advice item would be followed, the relatively high proportion of Q-nonagreements ($7$ of $20$ instances) might imply differences in knowledge across coders on how to map advice to lifecycle phases, or differing views of what comprises the \textit{design} phase. In this sense, we might hypothesize (untested) that Q-nonagreements at $Q_{11}$ were attributable to ambiguity in materials provided to coders (instructions, definitions) more than to unclear advice items. As $Q_{11}$ is/was a terminal internal tree node leading to two near-identical codes (one semantically a subset of the other), Q-nonagreements here (vs.\ at other questions) have relatively low impact on the final assigned code. In summary, $Q_{11}$ was visited relatively infrequently, had a relatively high proportion of nonagreements when visited by both coders, and provided marginal information even in the case of agreements; these reasons collectively motivate our simplification removing $Q_{11}$.

\textbf{Differing security experience across coders.} 
We consider C1 and C2 herein to be security experts, albeit one has seven to eight years more security experience. We use level of education, and time in security research, as rough measures of security expertise, and assume greater security knowledge given more experience in the domain. However, we did not formally assess coders' security expertise. Beyond analyzing reproducibility using additional coders (above), future work could investigate approaches to assess coder security expertise \cite{Giboney2016, Chi2006}. 

While difficult to gauge precisely, and not mentioned explicitly in the design of the coding tree, differences in experience may contribute to advice items and tree questions being interpreted and answered differently, increasing Q-nonagreements. An idea to address this (next point) is to revisit (refine the wording of) questions and annotations, aiming to reduce nonagreements due to different interpretations of ambiguous or unclear wording, and to individual coders relying on personal experience. 

\textbf{Clarity of terminology.} 
From Tables~\ref{tab:heatmap_2tag} and \ref{tab:heatmapSD_3tag}, we expect a likely contributing factor to outlier values  (the largest relative numbers of nonagreements) was coders C1 and C2 having different personal interpretations of terms in the key questions involved. If confirmed through further study, then the reproducibility of tagging might be improved by improving the clarity of terminology used in the tree questions and instructions to coders. From the observed outliers, we suggest exploring the sufficiency of definitions or explanations (for coders) of the following terms and phrases: 

-- \textit{unambiguous language, relatively focused} ($Q_1$, $M_1$)

-- \textit{outcome} ($Q_3$, $T$)

-- \textit{implied steps, explicit actions} ($Q_5$, $P_1$)  

-- \textit{security expert} ... and their capabilities/skillset ($Q_8$, $P_4$)
 
-- \textit{broad approach, security property} ($Q_{10}$, $P_2$)
 
\noindent While the existing question annotations \cite{Barrera2021a} provided to coders do add explanatory details, many of these terms remain without thorough explanations. Ideas to improve shared interpretations include refining the annotations to be more precise, with explicit definitions for more terms, and including concrete examples (perhaps multiple examples for some terms---e.g., in $Q_4$'s annotation describing \textit{techniques/mechanisms} and \textit{specific rules}). This then raises a methodological challenge: how to reliably deliver the refined annotations to coders, such that they are referenced and relied on. 

More generally, we reiterate the importance of unambiguous terminology, and the confusion that results from inconsistent use of poorly defined common terms (e.g., ``best practice''~\cite{Barrera2021a}). Examples of terminology recommended by authoritative sources include security glossaries from NIST \cite{NISTCSRC} and the IETF~\cite{Shirey1997}. However, aside from commonly used terminology often itself being ambiguous \cite{Barrera2021a}, customization of terminology to suit specific technical uses may be incompatible with standardization for widespread use.

\section{Related Work}

Here we discuss work related to qualitative dataset coding \cite{Corbin2008}, which is the basis of the SAcoding method~\cite{Barrera2021a}, to provide context on how other security researchers have used qualitative coding. For terminology and actionability of security advice, Barrera et al.~\cite{Barrera2021a} provide in-depth discussion of terminology used to establish and present security advice for IoT security practitioners. For practical use of the SAcoding method and an analysis of how security advice is constructed, Bellman and van Oorschot~\cite{bellman2022a} use the method to analyze and compare two sets of IoT security advice, and suggest desirable characteristics that contribute to actionable security advice.

McDonald et al. \cite{McDonald2019} note a surprising lack of details in the literature regarding descriptions of coding processes; this motivated in part our analysis herein. The following examples allow some comparison with our work, but represent only a small subset of qualitative coding papers in security-related research. 

Huaman et al.~\cite{Huaman2021} coded user feedback of password managers. After appearing to formally agree on a final codebook (versus describing agreement or calculating inter-rater reliability), they split their dataset among three coders who independently coded one third of the dataset each using their final codebook. 
Kang et al.~\cite{Kang2019} coded verbalized participant thoughts during drawing tasks about Internet-related tasks. They appear to use one coder for coding their dataset and include a second coder for 15\% of the dataset to calculate inter-coder agreement. Their process appears to use inter-coder agreement as a cross-check for their coding process, versus for establishing their codebook. 
Naiakshina et al.~\cite{Naiakshina2017} coded interview responses about how participants used secure password storage mechanisms. They used two coders to independently create codebooks for the full dataset and compared their codebooks using Cohen's Kappa (it is unclear how this was done). 
Krombholz et al.~\cite{Krombholz2017} coded verbalized participant thoughts during a system configuration task. They used two coders to independently establish codes, agree on a final codebook, code their dataset, and calculate inter-coder reliability (it is unclear if the calculation is on test sets or the final full coding). 
Ukrop et al.~\cite{Ukrop2019} coded interview responses from participants that had evaluated certificate validation status outputs (e.g., warnings, errors). They used two coders to independently establish initial codebooks, which were merged through discussion, and responses were re-coded using the final codebook. A third coder coded half the responses and inter-rater reliability was calculated between the initial two coders and the third coder. 
Ion et al.~\cite{ion_2015} compared expert and non expert security practices by manually coding open ended responses (from participants) to requests for experts to suggest security practices for non-experts. These practices were coded by two raters who together developed a codebook; no other details are given regarding codebook construction beyond obtaining a Cohen's Kappa of 0.77.

We note that the SAcoding method has not been used to code user-provided qualitative data (e.g., from participant interviews in user studies), although the datasets analyzed to date \cite{Barrera2021a, bellman2022a} were comprised of qualitative data (security advice items), and the analysis employs similar methods. The original design of the SAcoding tree~\cite{Barrera2021a} involved use of three coders on multiple test sets to iterate on a codebook, with a final coder mean pairwise agreement of 73\%\footnote{As noted earlier, this value was for tag agreements with the benefit of the $\pm 1$ rule based on a tag continuum~\cite{Barrera2021a}, and as such is not directly  comparable to the 46\% tag agreement reported in the present paper.} and ``substantial'' \cite{Landis1977} inter-rater reliability ($\kappa = 0.67$ mean; kappa values are not always given \cite{Huaman2021, Kang2019}). Rather than splitting the dataset across multiple coders (as in \cite{Huaman2021, Kang2019}), our analysis in the present paper involved two coders each independently tagging a full dataset (of 1013 items in our case), allowing comparison of full results. 

While not the primary focus of this paper,\footnote{
   Our analyses herein are perhaps most relevant for advice givers crafting advice for security practitioners/manufacturers rather than for end-users. 
} the SAcoding method can also be used to analyze security advice intended for other audiences, including end-users. We briefly mention select works covering user advice here. Redmiles et al.~\cite{redmiles_2016_ccs,Redmiles2016} identify various sources of advice and under what circumstances users accept or reject that advice. In subsequent work, Redmiles et al.~\cite{Redmiles2018} analyzed 1878 internet security advice documents and measured readability, finding that less than 25\% of analyzed advice would be understandable by a typical user. They also find that government and educational organizations tend to produce less readable advice than other sources. The SAcoding method considers the readability of advice in $Q_1$, where advice is tagged as non-actionable ($M_1$) if unclear or unfocused on a single topic (cf. Table~\ref{nowTable:treeQuestions}). More recently, Redmiles et al.~\cite{Redmiles2020} interviewed participants to estimate advice quality, including comprehensibility, perceived actionability, and perceived efficacy. The authors shared the codebook from Ion et al.~\cite{ion_2015}.

\section{Concluding Remarks}

Our comparison of a second coders' results to those a single coder in previous work \cite{Barrera2021a} provided substantial insight on the extent to which SAcoding results are reproducible across coders. We identified specific areas where challenges arose in reproducibly tagging a 1013-item security advice dataset. Even for the tag categories for which coders assigned a given tag to similar proportions of the dataset, detailed analysis revealed that the coders often failed to assign the identical tag to the same individual items, with the overall rate of tag agreement across the dataset being 46\%. A more positive result was the high agreement (80\% for SS-type comparisons, 87\% for SD-type) in identifying individual advice items as actionable vs.\ non-actionable. Our explanation for co-existence of these two apparently contradictory results is that when coders disagreed on tags for an item, in many cases the differing tags were nonetheless closely related, e.g., adjacent on the tag continuum of Barrera et al.~\cite{Barrera2021a}.

Despite the design intent of the SAcoding method to reduce   subjectivity in assigning tags to items, natural language descriptions of security advice are inherently imprecise. It is thus unsurprising that even using SAcoding (vs.\ directly assigning tags), the two coders tagged significant numbers of advice items differently. This leads to asking how nonagreements between coders might be further reduced. One idea that we have noted is to explore rewording or clarification of questions that resulted in relatively large numbers of Q-nonagreements, and to provide further explanatory materials and instructions to coders. 

We expect that if further studies were run, one with two coders of similar expert-level security experience and another with two of vastly different experience levels, fewer tag nonagreements would result within the first pairing.  This leads to asking: Is the goal of the SAcoding method to minimize nonagreements for coders of similar background, or between coders of vastly different background? Ideally, one might minimize nonagreements independent of coder experience---although we suggest that the SAcoding method itself is best limited to security experts meeting some lower bound of expertise (Barrera et al.~\cite{Barrera2021a} are silent on this). Beyond simply more coders, the effect of different sets of coders on SAcoding results and nonagreements remains a largely unexplored area. 

The SAcoding method and our analysis of it, including the introduction of novel techniques to explore question and tag nonagreements, such as tag-vs-tag tables, have allowed new measurements and insights on security advice. We hope that further research, including by others, will extend this work to broaden our understanding of the method's utility, and to refine its effectiveness. The general approach of the SAcoding method, and the construction and use of a coding tree for analysis of qualitative data, may also be of independent interest beyond the use case of analyzing security advice datasets. 

\textbf{Acknowledgements and Funding.} 
We thank anonymous referees for helpful comments. This work was supported by the Natural Sciences and Engineering Research Council of Canada (NSERC), which is acknowledged for Discovery Grants to the first and third authors, and for funding the third author as Canada Research Chair in Authentication and Computer Security. 

\bibliographystyle{ACM-Reference-Format}
\bibliography{bib}


\begin{thebibliography}{28}


\ifx \showCODEN    \undefined \def \showCODEN     #1{\unskip}     \fi
\ifx \showDOI      \undefined \def \showDOI       #1{#1}\fi
\ifx \showISBNx    \undefined \def \showISBNx     #1{\unskip}     \fi
\ifx \showISBNxiii \undefined \def \showISBNxiii  #1{\unskip}     \fi
\ifx \showISSN     \undefined \def \showISSN      #1{\unskip}     \fi
\ifx \showLCCN     \undefined \def \showLCCN      #1{\unskip}     \fi
\ifx \shownote     \undefined \def \shownote      #1{#1}          \fi
\ifx \showarticletitle \undefined \def \showarticletitle #1{#1}   \fi
\ifx \showURL      \undefined \def \showURL       {\relax}        \fi
\providecommand\bibfield[2]{#2}
\providecommand\bibinfo[2]{#2}
\providecommand\natexlab[1]{#1}
\providecommand\showeprint[2][]{arXiv:#2}

\bibitem[Alrawi et~al\mbox{.}(2019)]%
        {Alrawi2019}
\bibfield{author}{\bibinfo{person}{Omar Alrawi}, \bibinfo{person}{Chaz Lever},
  \bibinfo{person}{Manos Antonakakis}, {and} \bibinfo{person}{Fabian Monrose}.}
  \bibinfo{year}{2019}\natexlab{}.
\newblock \showarticletitle{{SoK: Security Evaluation of Home-Based IoT
  Deployments}}. In \bibinfo{booktitle}{\emph{{IEEE Symp.\ Security and
  Privacy}}}.
\newblock


\bibitem[{Association for Computing Machinery (ACM)}(2020)]%
        {ACMTerminology}
\bibfield{author}{\bibinfo{person}{{Association for Computing Machinery
  (ACM)}}.} \bibinfo{year}{{2020}}\natexlab{}.
\newblock \bibinfo{title}{{Artifact Review and Badging Version 1.1}}.
\newblock
  \bibinfo{howpublished}{\url{https://www.acm.org/publications/policies/artifact-review-and-badging-current}}.
\newblock


\bibitem[Barrera et~al\mbox{.}(2023)]%
        {Barrera2021a}
\bibfield{author}{\bibinfo{person}{David Barrera}, \bibinfo{person}{Christopher
  Bellman}, {and} \bibinfo{person}{Paul~C. van Oorschot}.}
  \bibinfo{year}{2023}\natexlab{}.
\newblock \showarticletitle{{Security Best Practices: A Critical Analysis Using
  IoT as a Case Study}}.
\newblock \bibinfo{journal}{\emph{ACM Trans. on Privacy and Security}}
  \bibinfo{volume}{26}, \bibinfo{number}{2} (\bibinfo{year}{2023}),
  \bibinfo{pages}{13:1--13:30}.
\newblock


\bibitem[Bellman(2022)]%
        {BellmanDataset}
\bibfield{author}{\bibinfo{person}{Christopher Bellman}.}
  \bibinfo{year}{2022}\natexlab{}.
\newblock \bibinfo{title}{{cb1013-dataset}}.
\newblock
  \bibinfo{howpublished}{\url{https://github.com/ChristopherBellman/SecurityAdvice/blob/main/cb1013-dataset-TOPS.json}}.
\newblock
\newblock
\shownote{Includes advice dataset and C1 tags}.


\bibitem[Bellman and Barrera(2022)]%
        {BellmanDatasetJCS}
\bibfield{author}{\bibinfo{person}{Christopher Bellman} {and}
  \bibinfo{person}{David Barrera}.} \bibinfo{year}{2022}\natexlab{}.
\newblock \bibinfo{title}{{cb1013-dataset-twocoder}}.
\newblock
  \bibinfo{howpublished}{\url{https://github.com/ChristopherBellman/SecurityAdvice/blob/main/cb1013-dataset-JCS.json}}.
\newblock
\newblock
\shownote{Includes advice dataset and C1, C2 tags}.


\bibitem[Bellman and van Oorschot(2023)]%
        {bellman2022a}
\bibfield{author}{\bibinfo{person}{Christopher Bellman} {and}
  \bibinfo{person}{Paul~C. van Oorschot}.} \bibinfo{year}{2023}\natexlab{}.
\newblock \showarticletitle{{Systematic Analysis and Comparison of Security
  Advice Datasets}}.
\newblock \bibinfo{journal}{\emph{Computers \& Security}}
  \bibinfo{volume}{124} (\bibinfo{year}{2023}), \bibinfo{pages}{102989}.
\newblock


\bibitem[Chi(2006)]%
        {Chi2006}
\bibfield{author}{\bibinfo{person}{Michelene T.~H. Chi}.}
  \bibinfo{year}{2006}\natexlab{}.
\newblock \showarticletitle{{Laboratory Methods for Assessing Experts' and
  Novices' Knowledge (Chapter 10)}}.
\newblock \bibinfo{journal}{\emph{{The Cambridge Handbook of Expertise and
  Expert Performance}}} (\bibinfo{year}{2006}).
\newblock


\bibitem[{Copper Horse Ltd.}(2019)]%
        {IoTSecMap}
\bibfield{author}{\bibinfo{person}{{Copper Horse Ltd.}}}
  \bibinfo{year}{2019}\natexlab{}.
\newblock \bibinfo{title}{{Mapping Security \& Privacy in the Internet of
  Things}}.
\newblock
  \bibinfo{howpublished}{\url{https://iotsecuritymapping.uk/wp-content/uploads/Mapping-of-Code-of-Practice-to-recommendations-and-standards_v3.json}}.
\newblock
\newblock
\shownote{Version 3 dataset}.


\bibitem[Corbin and Strauss(2008)]%
        {Corbin2008}
\bibfield{author}{\bibinfo{person}{Juliet Corbin} {and} \bibinfo{person}{Anselm
  Strauss}.} \bibinfo{year}{2008}\natexlab{}.
\newblock \bibinfo{booktitle}{\emph{{Basics of Qualitative Research: Techniques
  and Procedures for Developing Grounded Theory (third edition)}}}.
\newblock \bibinfo{publisher}{{SAGE Publications, Inc.}}
\newblock


\bibitem[{European Telecommunications Standards Institute (ETSI)}(2020)]%
        {ETSI2020}
\bibfield{author}{\bibinfo{person}{{European Telecommunications Standards
  Institute (ETSI)}}.} \bibinfo{year}{2020}\natexlab{}.
\newblock \bibinfo{title}{{CYBER; Cyber Security for Consumer Internet of
  Things: Baseline Requirements (ETSI EN 303 645)}}.
\newblock
  \bibinfo{howpublished}{\url{https://www.etsi.org/deliver/etsi_en/303600_303699/303645/02.01.01_60/en_303645v020101p.pdf}}.
\newblock


\bibitem[Giboney et~al\mbox{.}(2016)]%
        {Giboney2016}
\bibfield{author}{\bibinfo{person}{Justin~Scott Giboney},
  \bibinfo{person}{Jeffrey~Gainer Proudfoot}, \bibinfo{person}{Sanjay Goel},
  {and} \bibinfo{person}{Joseph~S. Valacich}.} \bibinfo{year}{2016}\natexlab{}.
\newblock \showarticletitle{{The Security Expertise Assessment Measure (SEAM):
  Developing a scale for hacker expertise}}.
\newblock \bibinfo{journal}{\emph{Computers \& Security}}  \bibinfo{volume}{60}
  (\bibinfo{year}{2016}), \bibinfo{pages}{37--51}.
\newblock


\bibitem[Huaman et~al\mbox{.}(2021)]%
        {Huaman2021}
\bibfield{author}{\bibinfo{person}{Nicholas Huaman}, \bibinfo{person}{Sabrina
  Amft}, \bibinfo{person}{Marten Oltrogge}, \bibinfo{person}{Yasemin Acar},
  {and} \bibinfo{person}{Sascha Fahl}.} \bibinfo{year}{2021}\natexlab{}.
\newblock \showarticletitle{{They Would do Better if They Worked Together: The
  Case of Interaction Problems Between Password Managers and Websites}}. In
  \bibinfo{booktitle}{\emph{IEEE Symp.\ Security and Privacy}}.
  \bibinfo{pages}{1626--1640}.
\newblock


\bibitem[Ion et~al\mbox{.}(2015)]%
        {ion_2015}
\bibfield{author}{\bibinfo{person}{Iulia Ion}, \bibinfo{person}{Rob Reeder},
  {and} \bibinfo{person}{Sunny Consolvo}.} \bibinfo{year}{2015}\natexlab{}.
\newblock \showarticletitle{"...{No} one {Can} {Hack} {My} {Mind}": {Comparing}
  {Expert} and {Non}-{Expert} {Security} {Practices}}. In
  \bibinfo{booktitle}{\emph{{USENIX} {SOUPS}}}.
\newblock


\bibitem[Kang et~al\mbox{.}(2019)]%
        {Kang2019}
\bibfield{author}{\bibinfo{person}{Ruogu Kang}, \bibinfo{person}{Laura
  Dabbish}, \bibinfo{person}{Nathaniel Fruchter}, {and} \bibinfo{person}{Sara
  Kiesler}.} \bibinfo{year}{2019}\natexlab{}.
\newblock \showarticletitle{{``My data just goes everywhere'': User Mental
  Models of the Internet and Implications for Privacy and Security}}. In
  \bibinfo{booktitle}{\emph{Symposium on Usable Privacy and Security (SOUPS)}}.
\newblock


\bibitem[Kolias et~al\mbox{.}(2017)]%
        {Kolias2017}
\bibfield{author}{\bibinfo{person}{Constantinos Kolias},
  \bibinfo{person}{Georgios Kambourakis}, \bibinfo{person}{Angelos Stavrou},
  {and} \bibinfo{person}{Jeffrey Voas}.} \bibinfo{year}{2017}\natexlab{}.
\newblock \showarticletitle{{DDoS in the IoT: Mirai and Other Botnets}}.
\newblock \bibinfo{journal}{\emph{Computer}} \bibinfo{volume}{50},
  \bibinfo{number}{7} (\bibinfo{year}{2017}), \bibinfo{pages}{80--84}.
\newblock


\bibitem[Krombholz et~al\mbox{.}(2017)]%
        {Krombholz2017}
\bibfield{author}{\bibinfo{person}{Katharina Krombholz},
  \bibinfo{person}{Wilfried Mayer}, \bibinfo{person}{Martin Schmiedecker},
  {and} \bibinfo{person}{Edgar Weippl}.} \bibinfo{year}{2017}\natexlab{}.
\newblock \showarticletitle{{``I Have No Idea What I'm Doing''---On the
  Usability of Deploying HTTPS}}. In \bibinfo{booktitle}{\emph{USENIX Security
  Symposium}}.
\newblock


\bibitem[Landis and Koch(1977)]%
        {Landis1977}
\bibfield{author}{\bibinfo{person}{J.~Richard Landis} {and}
  \bibinfo{person}{Gary~G. Koch}.} \bibinfo{year}{1977}\natexlab{}.
\newblock \showarticletitle{{The Measurement of Observer Agreement for
  Categorical Data}}.
\newblock \bibinfo{journal}{\emph{Biometrics}} \bibinfo{volume}{33},
  \bibinfo{number}{1} (\bibinfo{year}{1977}), \bibinfo{pages}{159--174}.
\newblock


\bibitem[McDonald et~al\mbox{.}(2019)]%
        {McDonald2019}
\bibfield{author}{\bibinfo{person}{Nora McDonald}, \bibinfo{person}{Sarita
  Schoenebeck}, {and} \bibinfo{person}{Andrea Forte}.}
  \bibinfo{year}{2019}\natexlab{}.
\newblock \showarticletitle{{Reliability and Inter-rater Reliability in
  Qualitative Research: Norms and Guidelines for CSCW and HCI Practice}}.
\newblock \bibinfo{journal}{\emph{Proc. ACM Hum.-Comput. Interact}}
  \bibinfo{volume}{3}, \bibinfo{number}{CSCW} (\bibinfo{date}{Nov}
  \bibinfo{year}{2019}), \bibinfo{pages}{1--23}.
\newblock


\bibitem[Naiakshina et~al\mbox{.}(2017)]%
        {Naiakshina2017}
\bibfield{author}{\bibinfo{person}{Alena Naiakshina},
  \bibinfo{person}{Anastasia Danilova}, \bibinfo{person}{Christian Tiefenau},
  \bibinfo{person}{Marco Herzog}, \bibinfo{person}{Sergej Dechand}, {and}
  \bibinfo{person}{M. Smith}.} \bibinfo{year}{2017}\natexlab{}.
\newblock \showarticletitle{{Why Do Developers Get Password Storage Wrong? A
  Qualitative Usability Study}}. In \bibinfo{booktitle}{\emph{ACM CCS}}.
\newblock


\bibitem[{NIST Computer Security Resource Center}(2022)]%
        {NISTCSRC}
\bibfield{author}{\bibinfo{person}{{NIST Computer Security Resource Center}}.}
  \bibinfo{year}{2022}\natexlab{}.
\newblock \bibinfo{title}{{Glossary}}.
\newblock
\newblock
\newblock
\shownote{Sept 20, 2022. \url{https://csrc.nist.gov/glossary}}.


\bibitem[Redmiles et~al\mbox{.}(2016a)]%
        {redmiles_2016_ccs}
\bibfield{author}{\bibinfo{person}{Elissa~M. Redmiles}, \bibinfo{person}{Sean
  Kross}, {and} \bibinfo{person}{Michelle~L. Mazurek}.}
  \bibinfo{year}{2016}\natexlab{a}.
\newblock \showarticletitle{How {I} {Learned} to be {Secure}: a
  {Census}-{Representative} {Survey} of {Security} {Advice} {Sources} and
  {Behavior}}. In \bibinfo{booktitle}{\emph{{ACM} {CCS}}}.
\newblock
\urldef\tempurl%
\url{https://doi.org/10.1145/2976749.2978307}
\showDOI{\tempurl}


\bibitem[Redmiles et~al\mbox{.}(2016b)]%
        {Redmiles2016}
\bibfield{author}{\bibinfo{person}{Elissa~M. Redmiles}, \bibinfo{person}{Amelia
  Malone}, {and} \bibinfo{person}{Michelle~L. Mazurek}.}
  \bibinfo{year}{2016}\natexlab{b}.
\newblock \showarticletitle{{I Think They're Trying to Tell Me Something:
  Advice Sources and Selection for Digital Security}}.
\newblock \bibinfo{journal}{\emph{IEEE Symp. on Security and Privacy}}
  (\bibinfo{year}{2016}), \bibinfo{pages}{272--288}.
\newblock


\bibitem[Redmiles et~al\mbox{.}(2018)]%
        {Redmiles2018}
\bibfield{author}{\bibinfo{person}{Elissa~M. Redmiles}, \bibinfo{person}{M.
  Morales}, \bibinfo{person}{Lisa Maszkiewicz}, \bibinfo{person}{R. Stevens},
  \bibinfo{person}{Everest Liu}, \bibinfo{person}{Dhruv Kuchhal}, {and}
  \bibinfo{person}{Michelle~L. Mazurek}.} \bibinfo{year}{2018}\natexlab{}.
\newblock \showarticletitle{{First Steps Toward Measuring the Readability of
  Security Advice}}. In \bibinfo{booktitle}{\emph{Workshop on Technology and
  Consumer Protection}}.
\newblock


\bibitem[Redmiles et~al\mbox{.}(2020)]%
        {Redmiles2020}
\bibfield{author}{\bibinfo{person}{Elissa~M. Redmiles}, \bibinfo{person}{Noel
  Warford}, \bibinfo{person}{Amritha Jayanti}, \bibinfo{person}{Aravind
  Koneru}, \bibinfo{person}{Sean Kross}, \bibinfo{person}{M. Morales},
  \bibinfo{person}{R. Stevens}, {and} \bibinfo{person}{Michelle~L. Mazurek}.}
  \bibinfo{year}{2020}\natexlab{}.
\newblock \showarticletitle{{A Comprehensive Quality Evaluation of Security and
  Privacy Advice on the Web}}. In \bibinfo{booktitle}{\emph{USENIX Security
  Symposium}}.
\newblock


\bibitem[Shirey(2007)]%
        {Shirey1997}
\bibfield{author}{\bibinfo{person}{Robert~W. Shirey}.}
  \bibinfo{year}{2007}\natexlab{}.
\newblock \bibinfo{title}{{RFC4949: Internet Security Glossary, Version 2}}.
\newblock
\newblock
\newblock
\shownote{{IETF}}.


\bibitem[{UK Department for Digital, Culture, Media \& Sport (DCMS)}(2018a)]%
        {DCMS2}
\bibfield{author}{\bibinfo{person}{{UK Department for Digital, Culture, Media
  \& Sport (DCMS)}}.} \bibinfo{year}{2018}\natexlab{a}.
\newblock \bibinfo{title}{{Code of Practice for Consumer IoT Security}}.
\newblock
  \bibinfo{howpublished}{\url{https://assets.publishing.service.gov.uk/government/uploads/system/uploads/attachment_data/file/773867/Code_of_Practice_for_Consumer_IoT_Security_October_2018.pdf}}.
\newblock


\bibitem[{UK Department for Digital, Culture, Media \& Sport (DCMS)}(2018b)]%
        {DCMS1}
\bibfield{author}{\bibinfo{person}{{UK Department for Digital, Culture, Media
  \& Sport (DCMS)}}.} \bibinfo{year}{2018}\natexlab{b}.
\newblock \bibinfo{title}{{Mapping of IoT Security Recommendations, Guidance
  and Standards to the UK's Code of Practice for Consumer IoT Security}}.
\newblock
\newblock
\newblock
\shownote{2018,
  \url{https://assets.publishing.service.gov.uk/government/uploads/system/uploads/attachment_data/file/774438/Mapping_of_IoT__Security_Recommendations_Guidance_and_Standards_to_CoP_Oct_2018.pdf}}.


\bibitem[Ukrop et~al\mbox{.}(2019)]%
        {Ukrop2019}
\bibfield{author}{\bibinfo{person}{Martin Ukrop}, \bibinfo{person}{Lydia
  Kraus}, \bibinfo{person}{Vashek Matyas}, {and} \bibinfo{person}{H.~A.~M.
  Wahsheh}.} \bibinfo{year}{2019}\natexlab{}.
\newblock \showarticletitle{{Will You Trust This TLS Certificate? Perceptions
  of People Working in IT}}. In \bibinfo{booktitle}{\emph{Annual Computer
  Security Applications Conference (ACSAC)}}.
\newblock


\end{thebibliography}

\end{document}